\documentclass[journal]{IEEEtran}

\usepackage{cite}
\ifCLASSINFOpdf
\else
\fi

\usepackage{graphicx}
\usepackage{amsmath}
\usepackage{mathtools}
\usepackage{commath}
\usepackage{comment}
\usepackage{enumerate}
\usepackage{booktabs}
\usepackage{multirow} 
\usepackage{epstopdf}
\usepackage{array}
\usepackage{wrapfig}
\usepackage{subcaption}

\newcommand{\innermid}{\nonscript\;\delimsize\vert\nonscript\;}
\newcommand{\activatebar}{%
  \begingroup\lccode`\~=`\|
  \lowercase{\endgroup\let~}\innermid 
  \mathcode`|=\string"8000
}

\usepackage{enumitem}
\usepackage{amsthm,amssymb}
\usepackage{setspace}
\newcommand{\subparagraph}{}
\usepackage[compact]{titlesec}
\usepackage[breaklinks=true]{hyperref}

\hypersetup{colorlinks=true,
            linkcolor=blue,urlcolor=blue,
            anchorcolor=blue,citecolor=blue}
            
\usepackage[anythingbreaks]{breakurl}           
 
\makeatletter
\g@addto@macro{\UrlBreaks}{\UrlOrds}
\makeatother

\newcommand\contentFigSz{0.195}

\begin{document}

\title{Towards Perceptually Optimized End-to-end Adaptive Video Streaming}

\author{Christos~G.~Bampis, Zhi Li, Ioannis Katsavounidis, Te-Yuan Huang, Chaitanya Ekanadham and Alan C. Bovik
\thanks{C. G. Bampis and A. C. Bovik are with the Department
of Electrical and Computer Engineering, University of Texas at Austin, Austin,
USA (e-mail: bampis@utexas.edu; bovik@ece.utexas.edu). Z. Li, I. Katsavounidis, TY. Huang and C. Ekanadham are with Netflix Inc. This work is supported by Netflix Inc.}
}

\maketitle

\begin{abstract}

Measuring Quality of Experience (QoE) and integrating these measurements into video streaming algorithms is a multi-faceted problem that fundamentally requires the design of comprehensive subjective QoE databases and metrics. To achieve this goal, we have recently designed the LIVE-NFLX-II database, a highly-realistic database which contains subjective QoE responses to various design dimensions, such as bitrate adaptation algorithms, network conditions and video content. Our database builds on recent advancements in content-adaptive encoding and incorporates actual network traces to capture realistic network variations on the client device. Using our database, we study the effects of multiple streaming dimensions on user experience and evaluate video quality and quality of experience models. We believe that the tools introduced here will help inspire further progress on the development of \textit{perceptually-optimized} client adaptation and video streaming strategies. The database is publicly available at \url{http://live.ece.utexas.edu/research/LIVE_NFLX_II/live_nflx_plus.html}.

\end{abstract}


\IEEEpeerreviewmaketitle

\section{Introduction}

\IEEEPARstart{V}\ \hspace{-1.1mm}ideo traffic from content delivery networks is expected to occupy 71\% of all consumed bandwidth by 2021\cite{cisco}. The need for more bandwidth and network resources is largely fueled by large scale video streaming applications and by increasing consumer demand for better quality videos and larger display sizes. Nevertheless, the available bandwidth is sometimes volatile and/or limited, in particular on mobile networks.

HTTP-based adaptive video streaming (HAS) is becoming the \textit{de facto} standard for modern video streaming services, such as Netflix and YouTube. The main idea behind HAS is to encode video content into multiple streams of various bitrate and quality levels, and to allow for client-driven stream selection to meet the time-varying network bandwidth. Under this setting, the client device is responsible for deciding on the bitrate/quality level of the video chunk to be played next. These client decisions are usually based on past network throughput values, future throughput estimates and other client-related information, such as the device buffer status \cite{huang2015buffer}.

In HAS, TCP is used as the transfer protocol; hence packet loss is not an issue \cite{techblog}. Nevertheless, depending on the available bandwidth, client devices may adapt to different quality levels and hence users may suffer from compression/scaling artifacts and rebuffering. When the available bandwidth drops, a client may use a higher compression ratio and/or a lower encoding resolution to reduce the video bitrate, leading to compression and scaling artifacts, respectively \cite{techblog}. Compression artifacts are usually perceived as blocky regions or local flicker, while scaling artifacts are presented as visible blocking, blurring and/or halo artifacts. All these artifacts are spatial, i.e., the temporal aspect of video playback is uninterrupted. On the other hand, if the throughput reaches a very low value and the buffer is emptied, a client must pause its video playback (video rebuffering), wait for the network to recover, and fill the buffer with video data before resuming play.

These video impairments can adversely affect user Quality of Experience (QoE), i.e., the overall level of user satisfaction \cite{brunnstrom2013qualinet} while viewing streaming content. Being able to predict QoE, and act upon those predictions, is important for improving the overall viewer experience. Towards this goal, we can design algorithms to optimize the QoE while effectively utilizing the available bandwidth and reducing operational costs. Modeling QoE is a difficult task, since QoE is affected by many complex and sometimes inaccessible factors, while obtaining ground truth QoE data that reflects these many factors is difficult.

Understanding and predicting QoE for adaptive video streaming is an emerging research area \cite{garcia2014quality,seufert2015survey,tavakoli2016perceptual,tavakoli2015quality,6918432,duanmu2016sqi,bampis2017study,7931662,balachandran2013developing,hossfeld2012initial,VATL,mok2011measuring}. Existing QoE studies do not fully capture important aspects of practical video streaming systems, e.g., they do not incorporate actual network measurements and client adaptive bitrate adaptation (ABR) algorithms. To this end, we built the LIVE-NFLX-II database, a large subjective QoE database that integrates perceptual video coding and quality assessment, using real measurements of network and buffer conditions, and client-based adaptation. 

\begin{figure*}[tp]
\centerline{
\includegraphics[width=1.4\columnwidth]{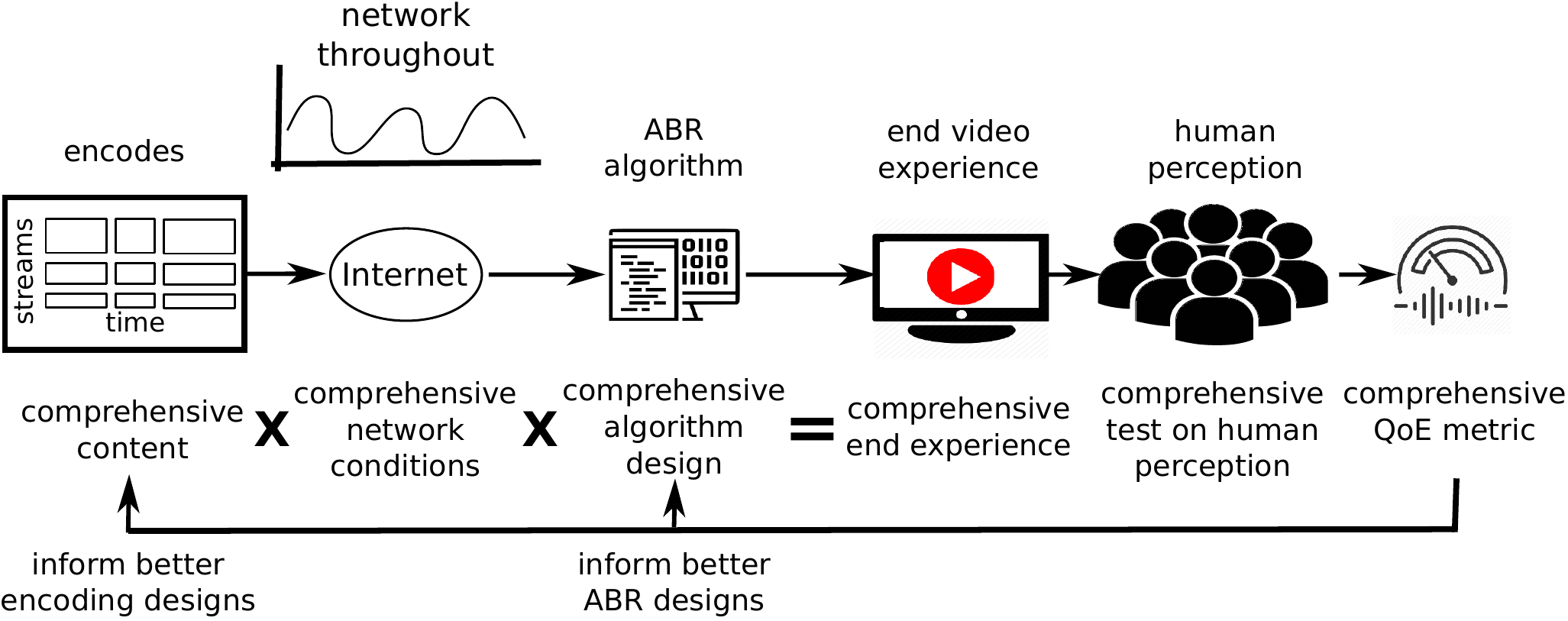}
}
\caption{An overview diagram of this work.}
\label{comprehensive}
\end{figure*}

Figure \ref{comprehensive} depicts a roadmap of our work. We first generate a large set of end-user experiences on top of a comprehensive collection of video contents, network conditions and ABR algorithms. Then, we conduct a comprehensive test on human perception to build enhanced QoE metrics. Ultimately, these metrics can be used to inform better ABR algorithms or encoding strategies.

A unique characteristic of the subjective database presented herein is that we incorporate recent developments in large-scale video encoding and ABR algorithms. To generate video encodes, we make use of an encoding optimization framework \cite{DO_techblog} that selects encoding parameters on a per-shot basis, guided by a state-of-the-art video quality assessment algorithm (VMAF) \cite{techblog}.

To model video streaming, we use actual network measurements and a pragmatic client buffer simulator, rather than just simplistic network and buffer occupancy models. Given the plethora of network traces and ABR algorithms, the database captures multiple streaming adaptation aspects, such as video quality fluctuations, rebuffering events of varying durations and occurrences, spatial resolution changes, and video content types. The subjective data consists of both retrospective and continuous-time scores, which makes it ideal for training various QoE models. Lastly, the video database is considerably larger than other public-domain video QoE databases \cite{chen2014modeling,duanmu2016sqi,ghadiyaram2017subjective}.

The main observations from the collected data can be summarized as follows. A better bandwidth prediction model can improve most of the objective streaming metrics, such as the playout bitrate and the number and duration of rebuffers. Meanwhile, the start-up phase is the most challenging part of a session for all ABR algorithms. This is because ABR algorithms have not built up the video buffer and hence network variations can easily impact QoE negatively. While this observation is in line with previous studies \cite{huang2015buffer}, we take this observation a step further. Our continuous data shows that humans can perceive these differences during start-up, even if they are forgiving and/or forgetful when a retrospective score is recorded. These observations highlight the importance of temporal studies of QoE, especially during start-up, on practical video streaming applications.

From a QoE model development perspective, we trained a retrospective QoE predictor on the dataset and determined that QoE predictions were mostly influenced by average video quality, followed by rebuffering duration. Lastly, we evaluated two state-of-the-art continuous-time QoE prediction models and found that they still fall short in their ability to capture trends in human responses, hence progress in this field would benefit from better models of human responses to these temporal phenomena.

We hope that our research efforts will help motivate and inspire further research on perceptually-optimized video streaming. The rest of this paper is organized as follows. Section \ref{prev_works} gives an overview of previous QoE studies and their shortcomings. Section \ref{recreating} discusses the comprehensive nature of the streaming database and streaming pipeline model we developed. In Section \ref{subjective_test}, the subjective testing procedure is discussed and an objective analysis of the database is presented in Section \ref{objective_analysis}. Following that, Section \ref{subjective_analysis} studies the collected human opinion scores. Section \ref{objective_VQA_QoE} evaluates video quality assessment (VQA) and QoE prediction models in the new dataset while Section \ref{the_end} concludes with future work.

\section{Related Work}
\label{prev_works}

Designing subjective databases to better understand human video perception has received great attention in the past. Many databases have been designed towards advancing progress on the more general problem of video quality \cite{moorthy2012video,de2010h,vu2014vis3,vqeg_hd3,hosu2017konstanz,winkler2012analysis} and streaming \cite{hossfeld2012initial,chen2014modeling,thang2014evaluation,rainer2014quality,6918432,tavakoli2015quality,tavakoli2016perceptual,duanmu2016sqi,bampis2017study, sogaard2017subjective,lin2015mcl,yeganeh2017joint}. There are also two valuable survey papers in the field in \cite{garcia2014quality,seufert2015survey}. Here we give only a brief overview to elucidate important shortcomings of previous studies, and to suggest improvements. 

The time-varying quality of long HTTP streams was investigated in \cite{chen2014modeling}, without considering rebuffering events and/or client ABR algorithms. A crowdsourcing experimental comparison among three representative HTTP-based clients was carried out in \cite{rainer2014quality}, but only one video content was used and continuous QoE was not investigated. In \cite{duanmu2016sqi}, the effects of rebuffering and quality changes were jointly considered, but the videos used were of short duration and the generated distortions did not simulate any actual client adaptation strategy. More recently, in \cite{duanmu2018quality}, the QoE behavior of various ABR algorithms was evaluated, but continuous QoE effects were not investigated. A simplistic buffer and network approach was derived in \cite{bampis2017study} to study the trade-offs between compression and rebuffering, but only eight distortions were generated and the database is not available in its entirety. Further, it was quite common for the aforementioned experiments to use a fixed bitrate ladder, without considering content-aware encoding strategies which are gaining popularity.

To summarize the main shortcomings of these previous studies, we believe that they do not study all of the multiple important dimensions in the client adaptation space, they do not use actual network measurements or a buffer occupancy model to depict a realistic streaming scenario, they are small or moderate in size, and they are not always publicly available. Table \ref{differences} demonstrates these shortcomings and highlights the contribution of the new database.

\begin{table*}[htp]
\caption{High-level comparison with other relevant video streaming subjective studies.}
\centering
\scalebox{1}{
\begin{tabular}{| c | c | c | c | c | c | c | c | c | c | c | c |}
\hline
Description & \cite{rainer2014quality} & \cite{6918432} & \cite{tavakoli2016perceptual} & \cite{yeganeh2017joint} & \cite{lin2015mcl} & \cite{chen2014modeling} & \cite{duanmu2016sqi} & \cite{ghadiyaram2017subjective} & \cite{bampis2017study} & \cite{duanmu2018quality} & LIVE-NFLX-II \\ \hline
client adaptation & X & & & & & & & & & X & X \\ \hline
continuous QoE & & X & & & & X & & X & X & & X \\ \hline
actual network traces & X & & & & & & & & & & X \\ \hline
buffer model & X & & & & & & & & X & X & X \\ \hline
public & & & & & X & X & X & X & & X & X \\ \hline
$>$ 400 test videos & & & & & & & & & & X & X \\ \hline
$>$ 60 subjects & X & & X & & & & & & & & X \\ \hline
rebuffering + quality & X & X & & X & & & X & & X & X & X \\ \hline
content-based encoding & & & X & & X & & & & & & X \\ \hline
\end{tabular}}
\label{differences}
\end{table*}

Here we present our efforts to advance the field of \textit{perceptually-optimized} adaptive video streaming by designing a new and unique QoE database, whereby perceptual video quality principles are injected into various stages of a modern streaming system: encoding, quality monitoring and client adaptation. We hope that these efforts will help pave the way for the development of optimized perceptual streaming systems.

\section{Recreating A Comprehensive End-user Experience}
\label{recreating}

\subsection{Overview of the Streaming System}

\begin{figure}[tp]
\centerline{
\includegraphics[width=0.7\columnwidth]{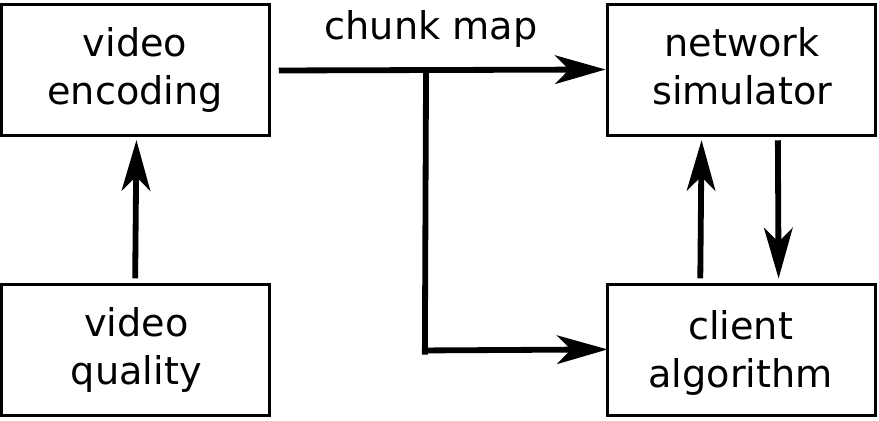}
}
\caption{Overview of modules in the proposed adaptive streaming pipeline to generate video streams for the subjective study.}
\label{pipeline_overview}
\end{figure}

To overcome the limitations of previous QoE studies, we built our database based on a highly realistic adaptive streaming pipeline model, which comprises four main modules, as shown in Fig. \ref{pipeline_overview}. The term ``module" describes a set of operations that are being carried out for video encoding (encoding module), video quality calculation (video quality module), network transmission (network module) and client-based video playout (client module). 

The encoding module constructs a content-driven bitrate ladder which is then fed into the Dynamic Optimizer (DO) \cite{DO_techblog}: a state-of-the-art encoding optimization approach, which determines the encoding parameters (encoding resolution and Quantization Parameter - QP) that produces encodes of optimized quality. The video quality module performs VMAF \cite{techblog} quality measurements that drive the encoding and client modules. The VMAF quality measurements are stored in a chunk map and made available on the client side for client bitrate adaptation. A chunk map contains information about every encoded video segment, see Fig. \ref{encoding_chunk_map} for more details. The network module incorporates the selected network traces and is responsible for communication between the encoding, video quality and client modules. The client module is responsible for requesting the next chunk to be played. Sections \ref{encoding_module_appendix} and \ref{video_quality_module_appendix} provide more details on the encoding and video quality modules respectively, while Section \ref{pieces_together_appendix} further discusses the streaming pipeline model we built.

This streaming model allowed us to recreate a comprehensive end-user experience by focusing on three streaming dimensions: encoding, network throughout and the choice of ABR algorithm. To study each of these dimensions, we incorporated 15 video contents, 7 actual network traces and 4 adaptation algorithms, yielding 420 video streams in total. Next, we explore the diverse characteristics of each dimension with the overarching goal of recreating a comprehensive end-user experience.

\subsection{Video Encoding}

\begin{figure*}
\centering
\subcaptionbox*{AirShow}{\includegraphics[width=\contentFigSz\textwidth]{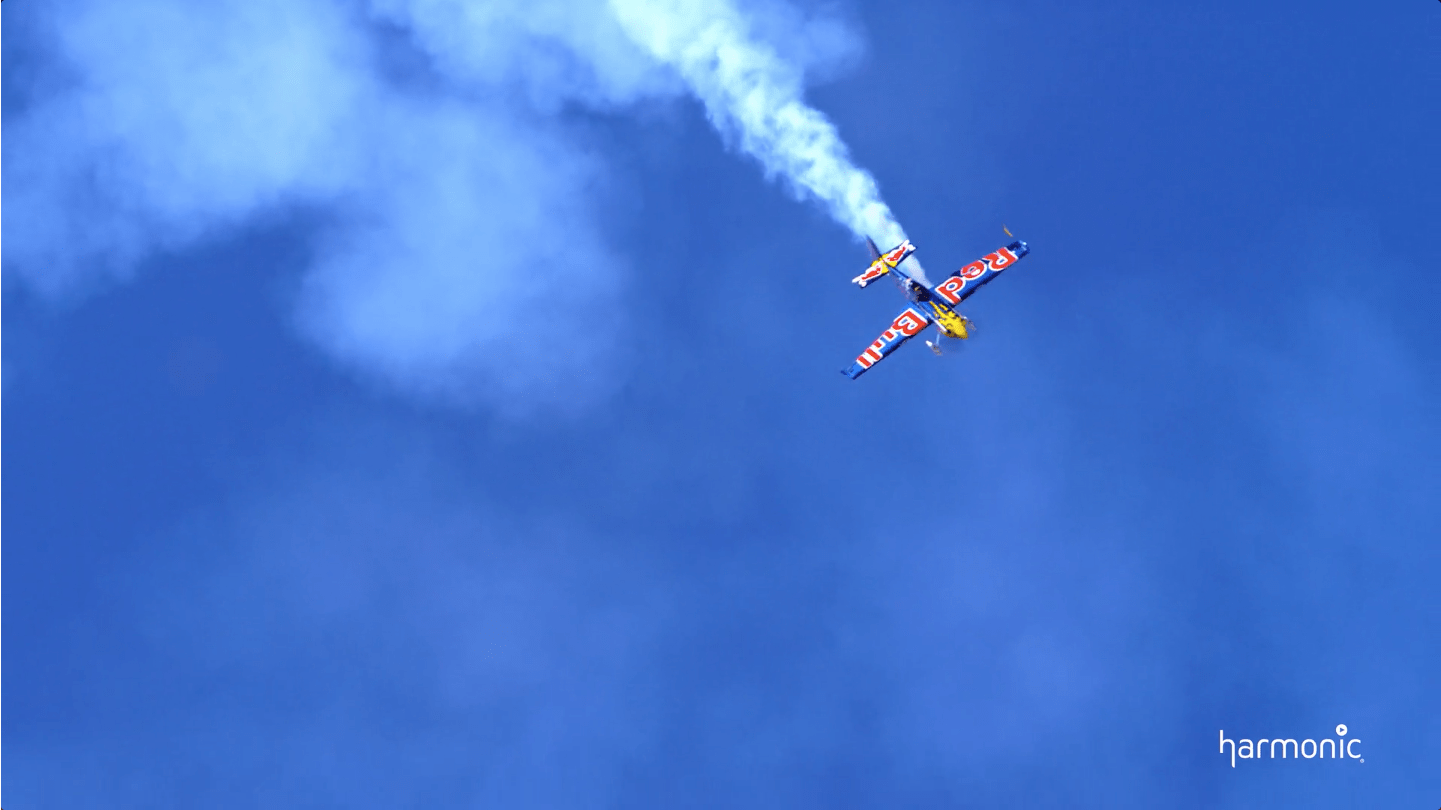}}\hfill
\subcaptionbox*{AsianFusion}{\includegraphics[width=\contentFigSz\textwidth]{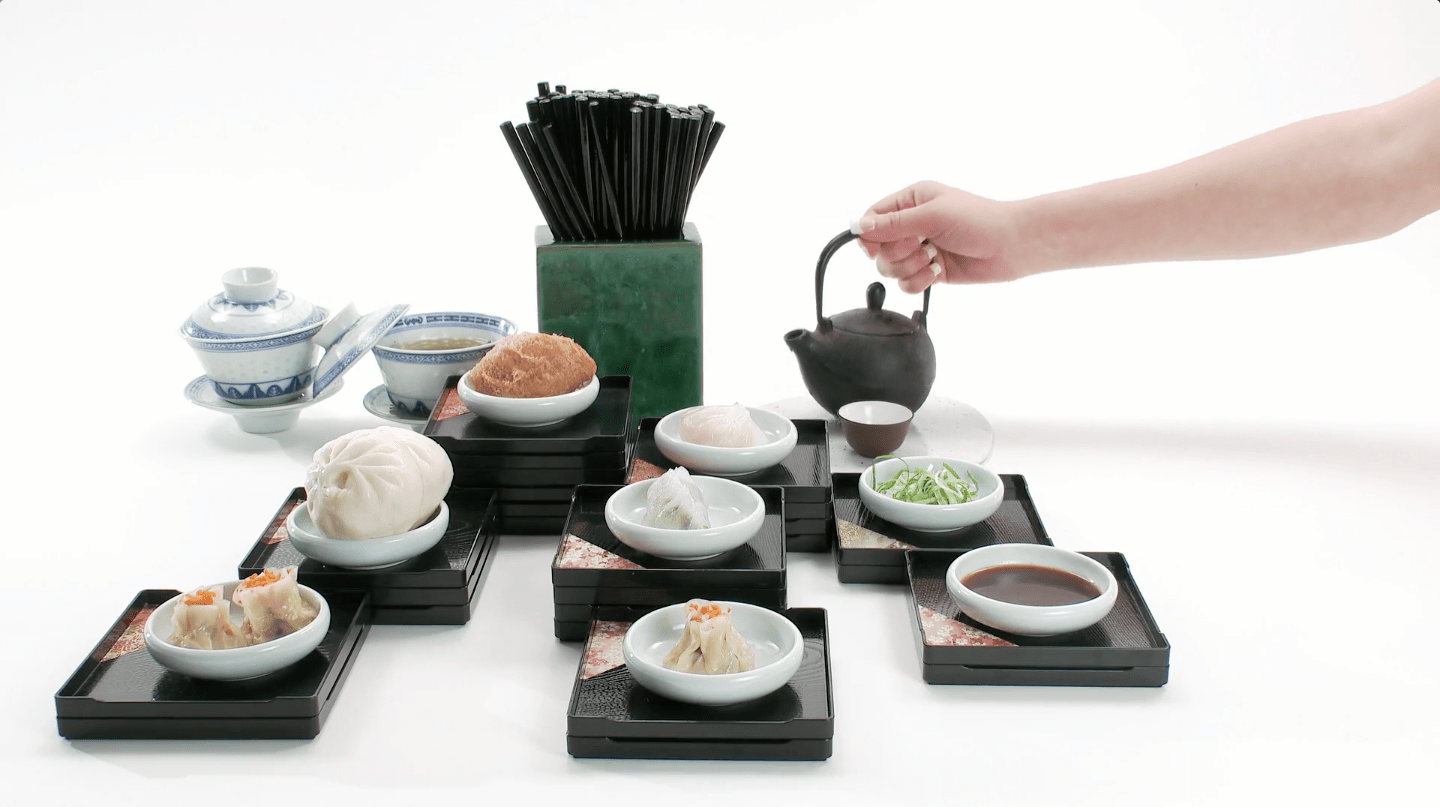}}\hfill
\subcaptionbox*{Chimera1102353}{\includegraphics[width=\contentFigSz\textwidth]{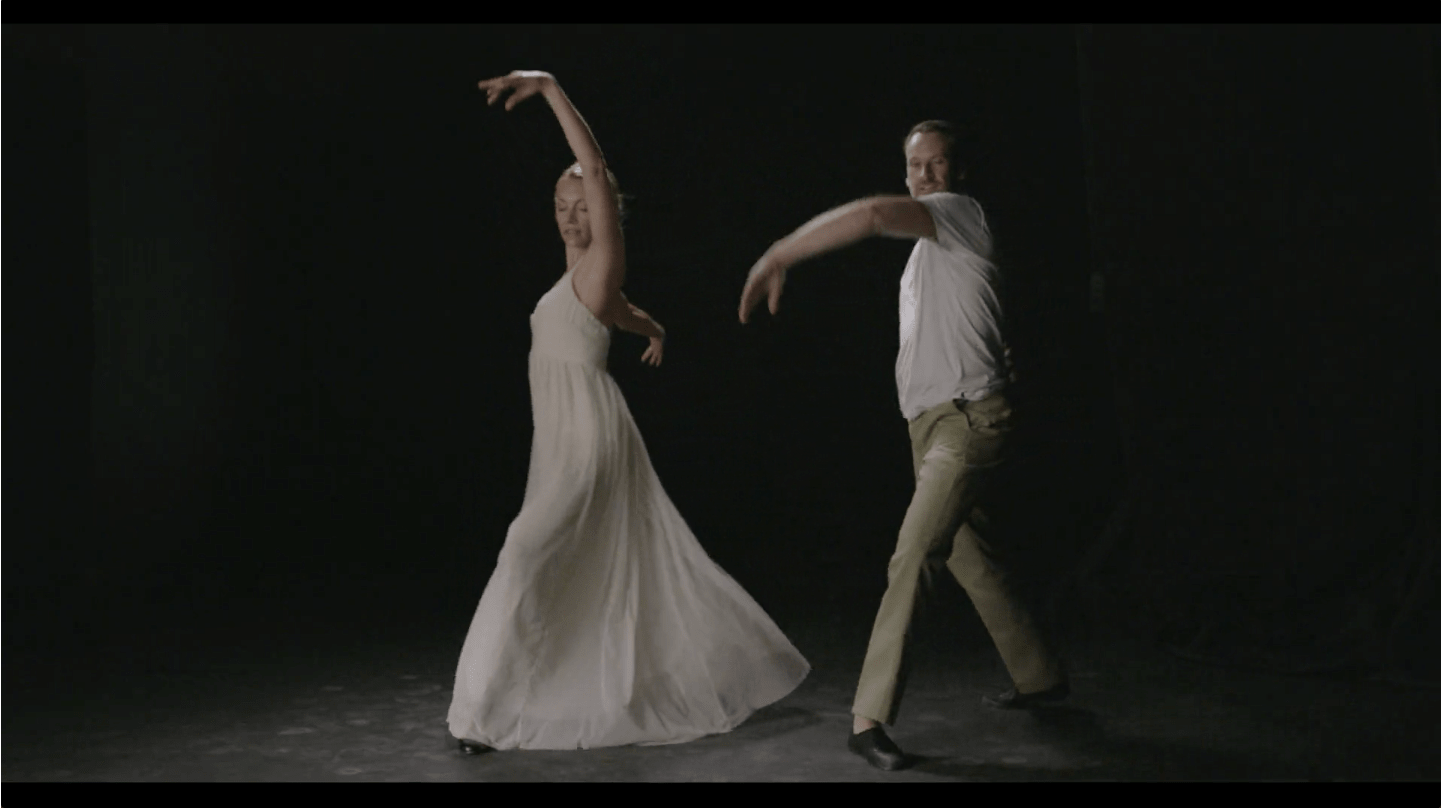}}\hfill
\subcaptionbox*{Chimera1102347}{\includegraphics[width=\contentFigSz\textwidth]{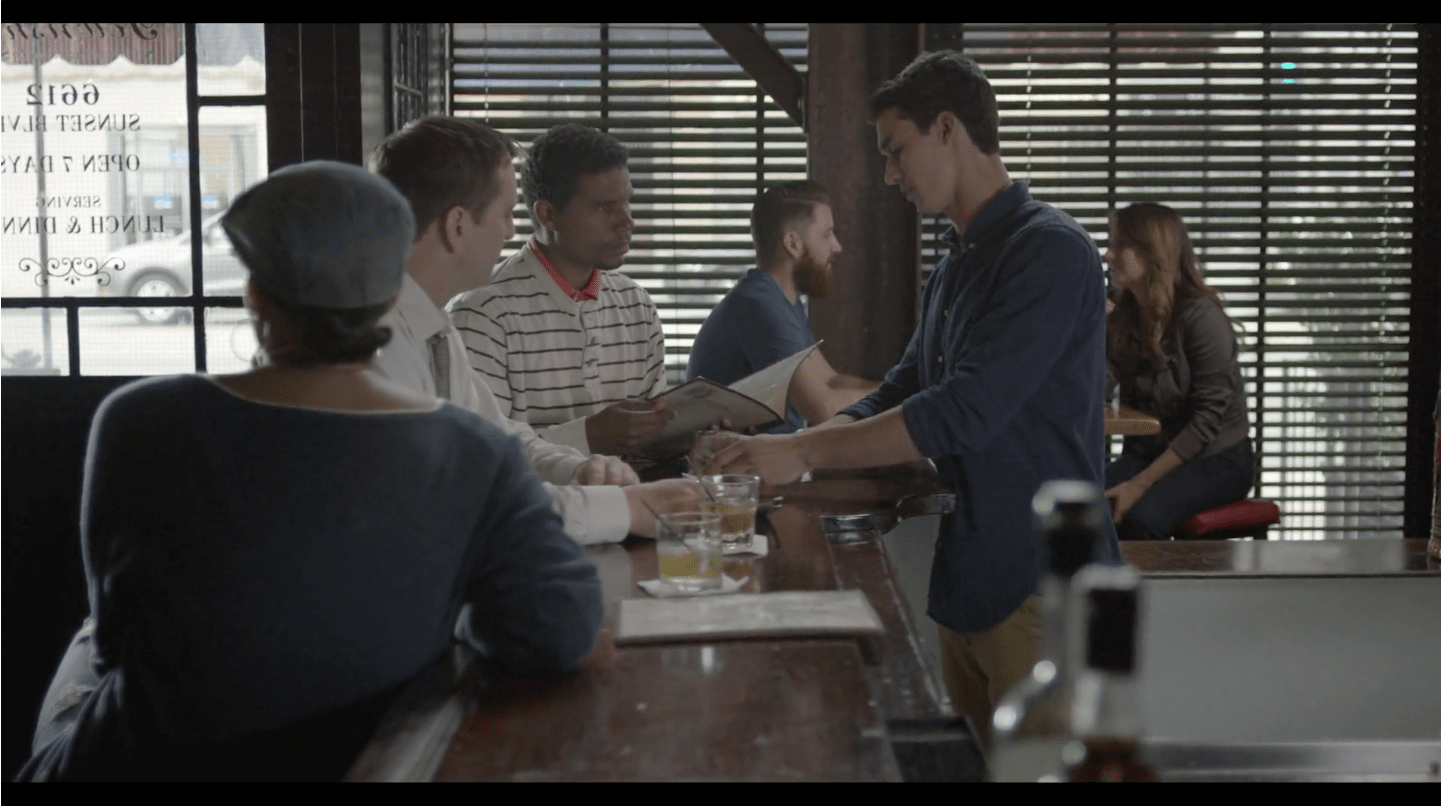}}\hfill
\subcaptionbox*{CosmosLaundromat}{\includegraphics[width=\contentFigSz\textwidth]{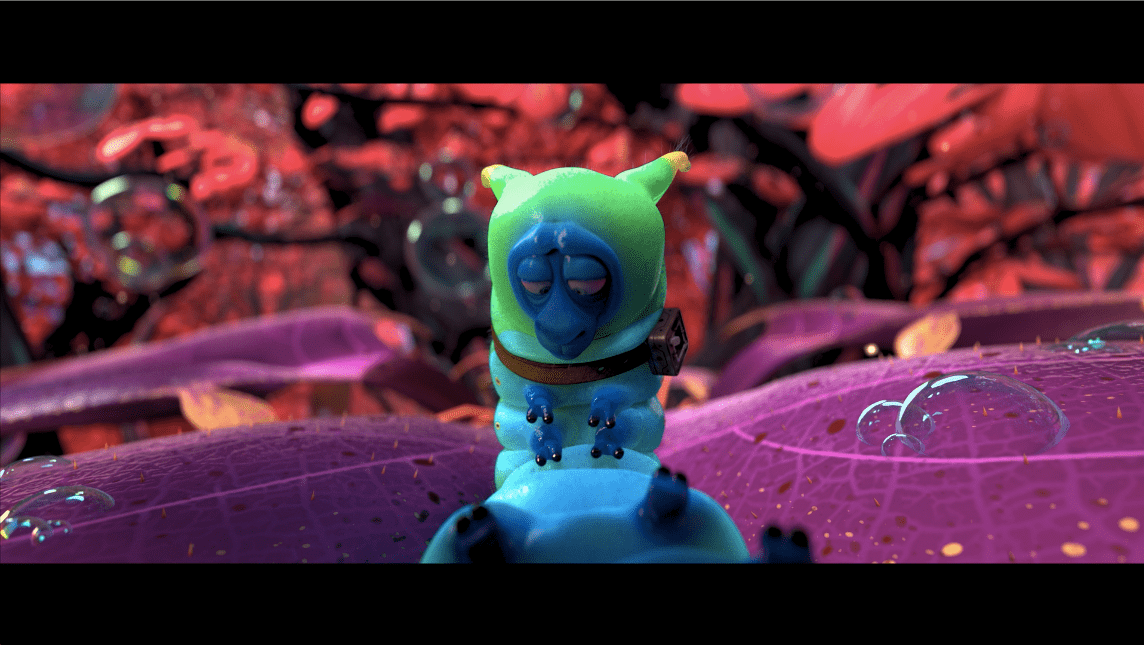}}\hfill
\vspace{1mm}
\subcaptionbox*{ElFuenteDance}{\includegraphics[width=\contentFigSz\textwidth]{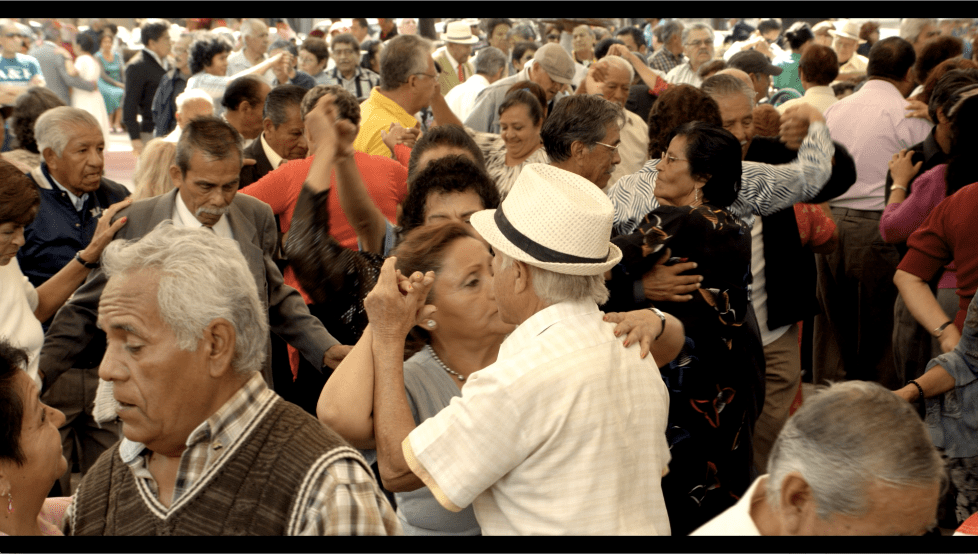}}\hfill
\subcaptionbox*{ElFuenteMask}{\includegraphics[width=\contentFigSz\textwidth]{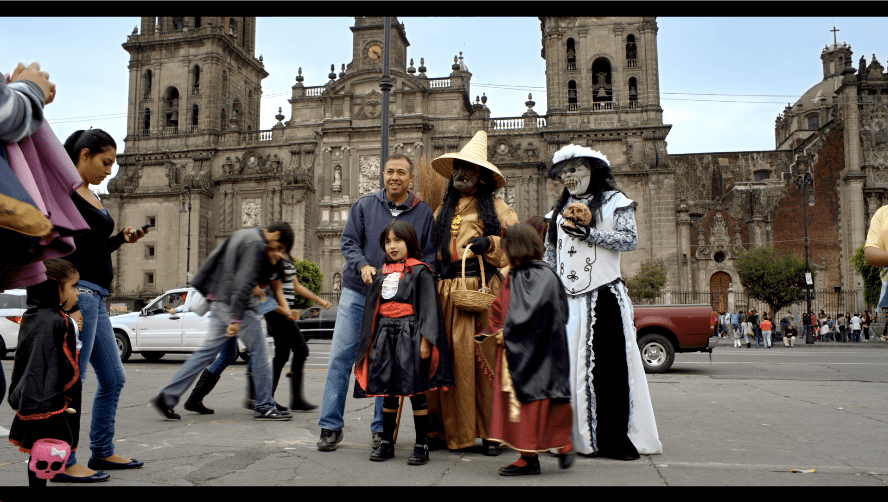}}\hfill
\subcaptionbox*{GTA}{\includegraphics[width=\contentFigSz\textwidth]{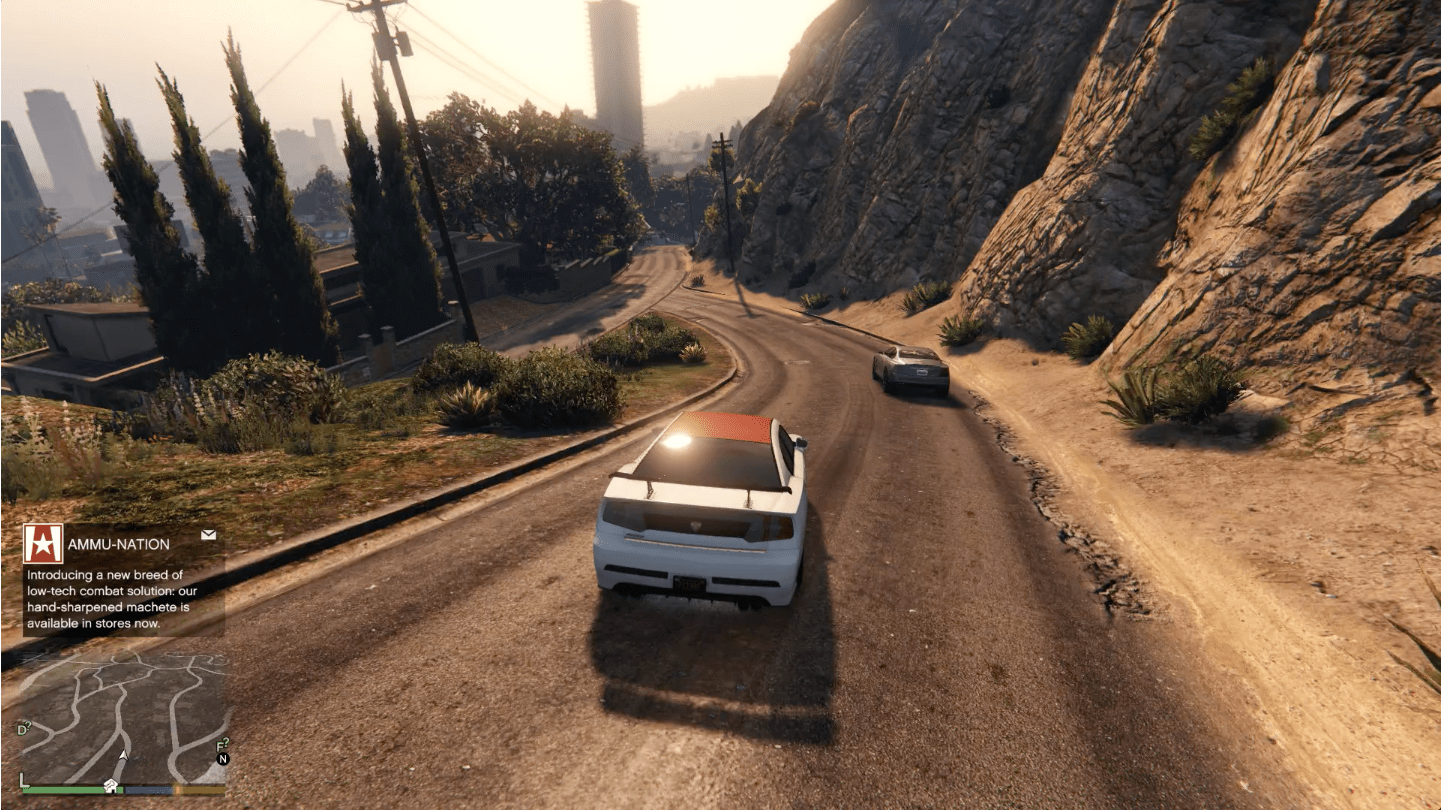}}\hfill
\subcaptionbox*{MeridianConversation}{\includegraphics[width=\contentFigSz\textwidth]{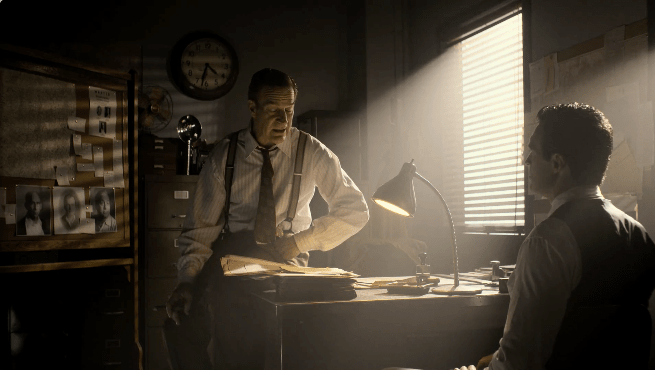}}\hfill
\subcaptionbox*{MeridianDriving}{\includegraphics[width=\contentFigSz\textwidth]{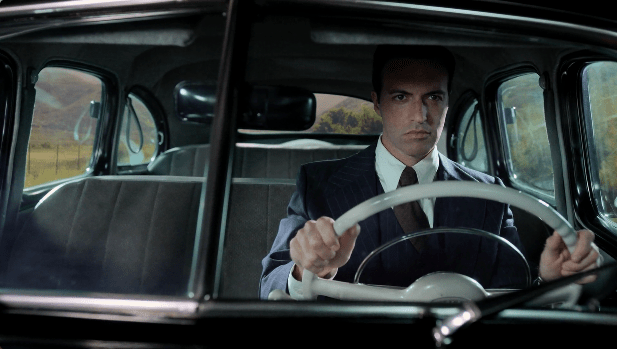}}\hfill
\vspace{1mm}
\subcaptionbox*{Skateboarding}{\includegraphics[width=\contentFigSz\textwidth]{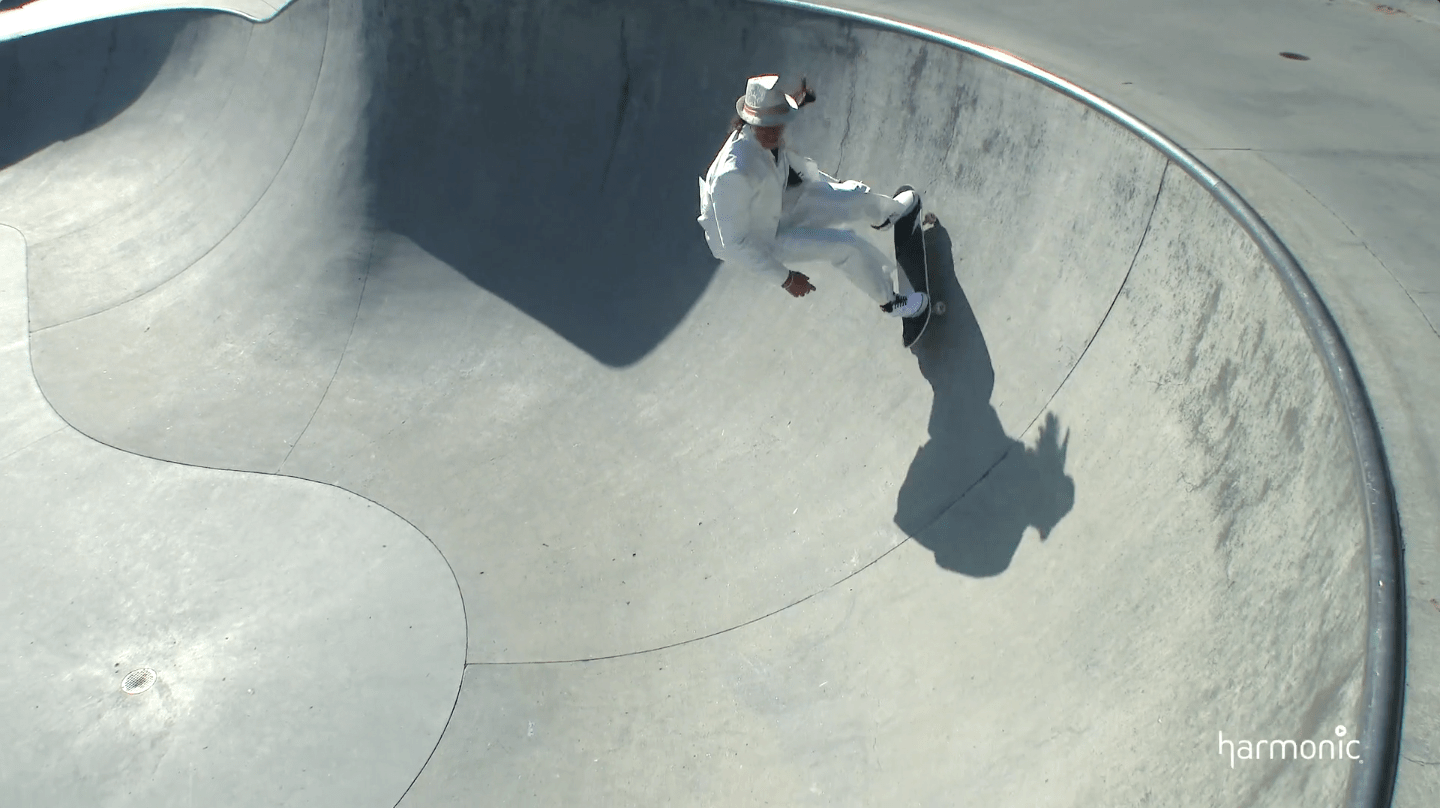}}\hfill
\subcaptionbox*{Soccer}{\includegraphics[width=\contentFigSz\textwidth]{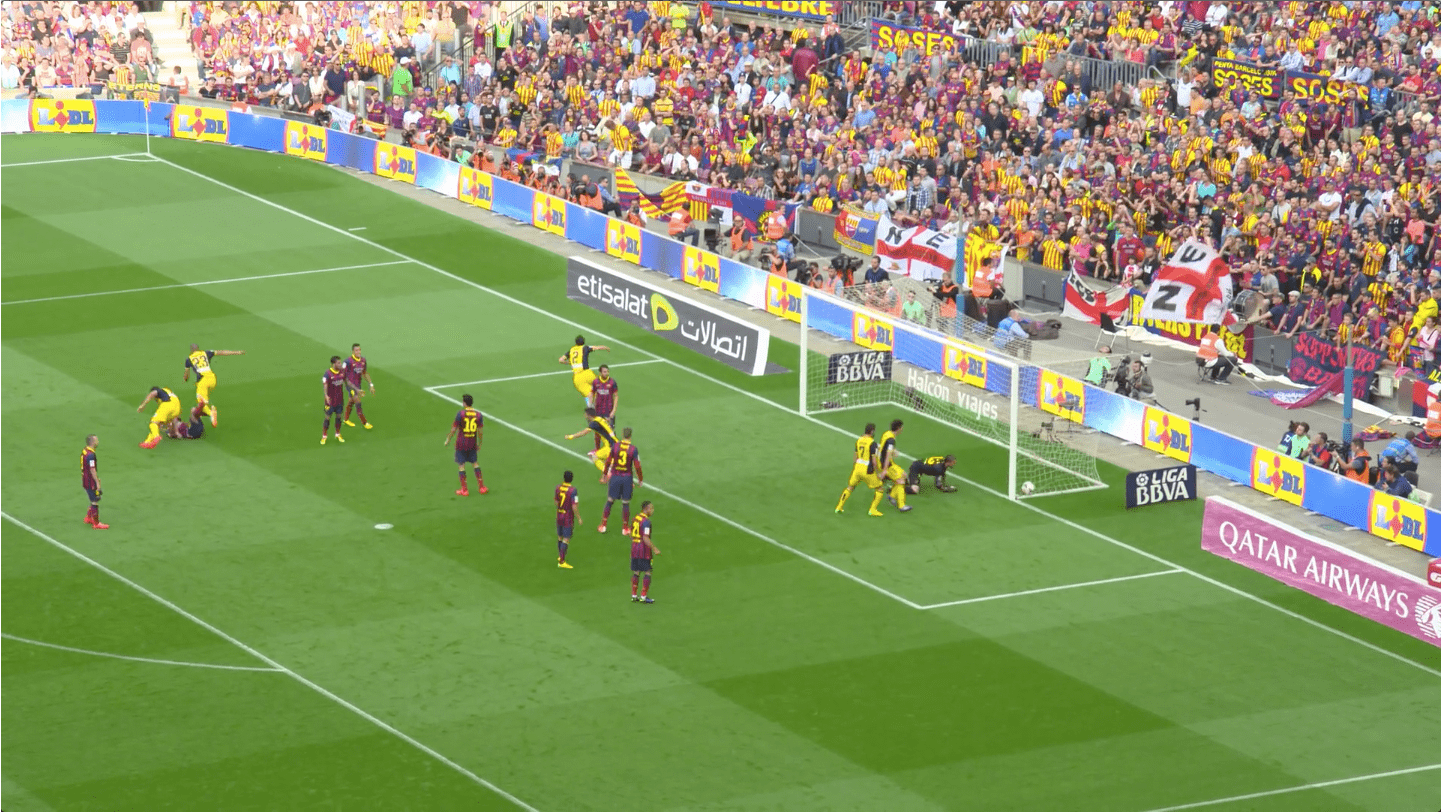}}\hfill
\subcaptionbox*{Sparks}{\includegraphics[width=\contentFigSz\textwidth]{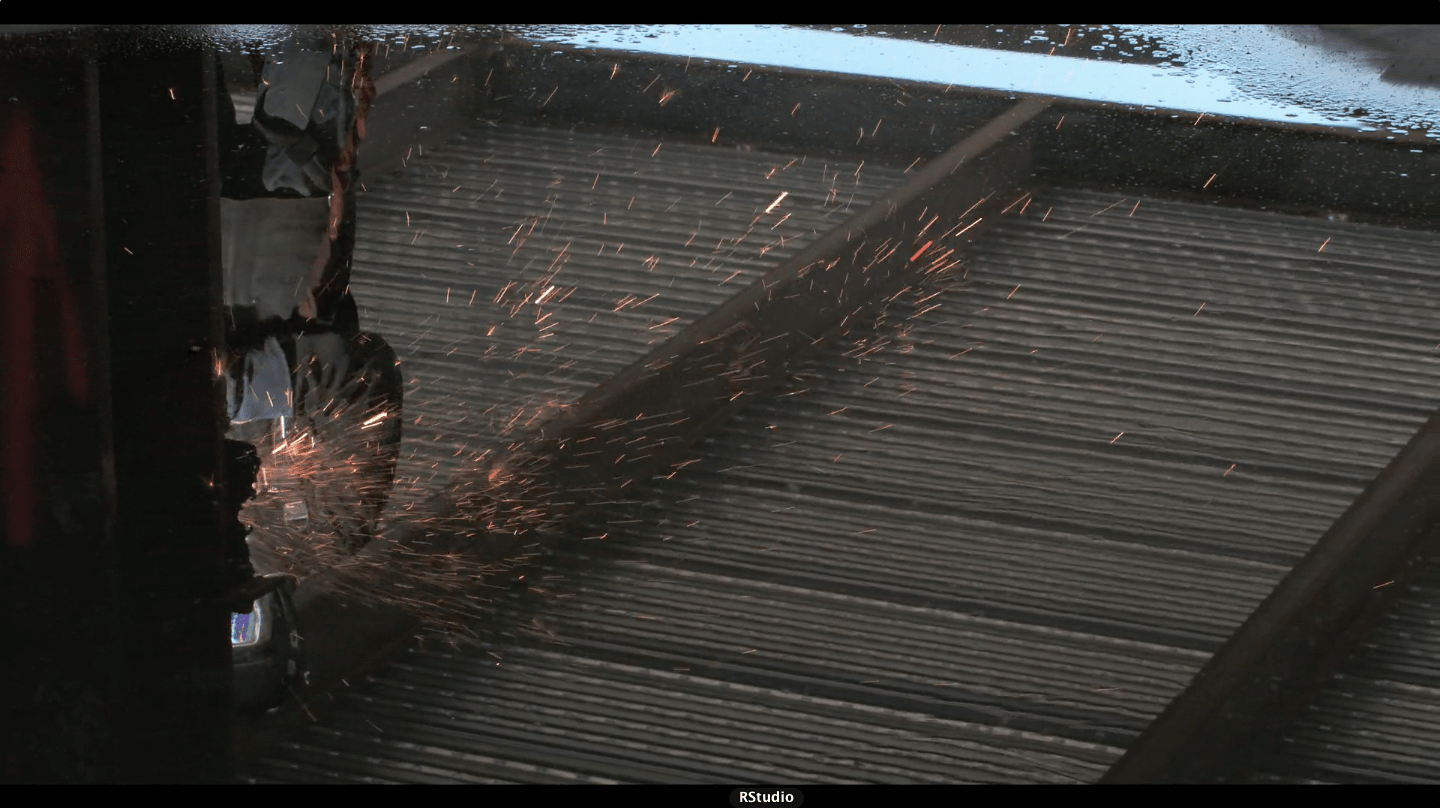}}\hfill
\subcaptionbox*{TearsOfSteelRobot}{\includegraphics[width=\contentFigSz\textwidth]{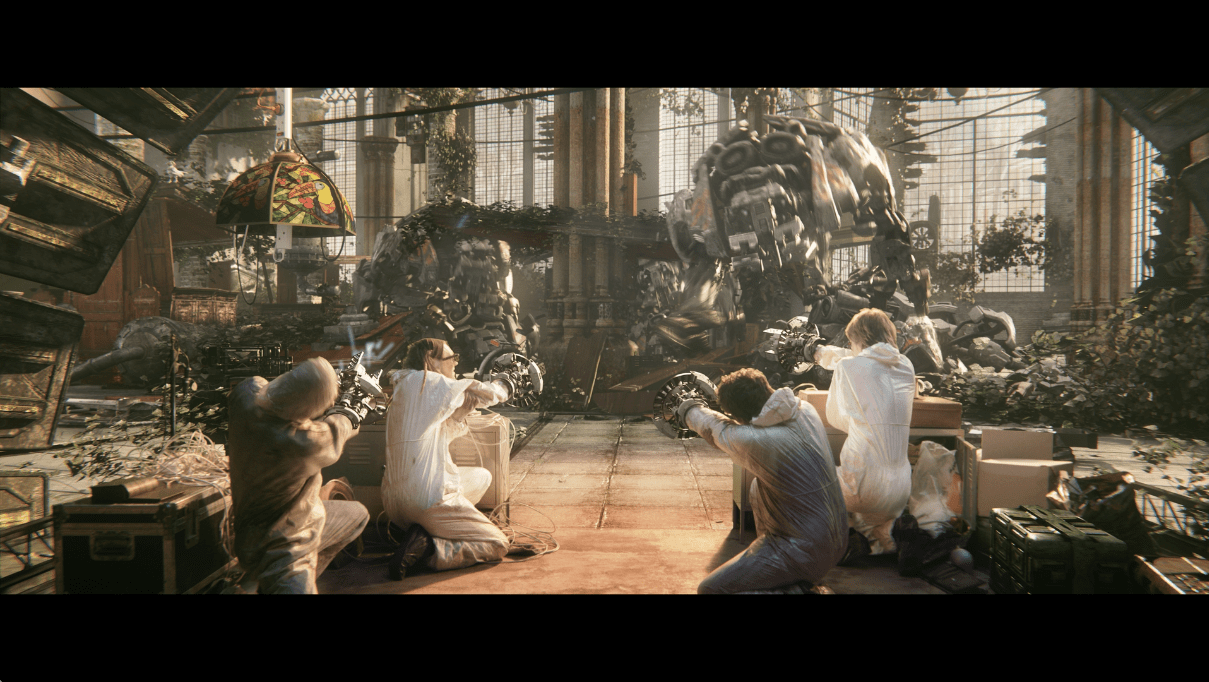}}\hfill
\subcaptionbox*{TearsOfSteelStatic}{\includegraphics[width=\contentFigSz\textwidth]{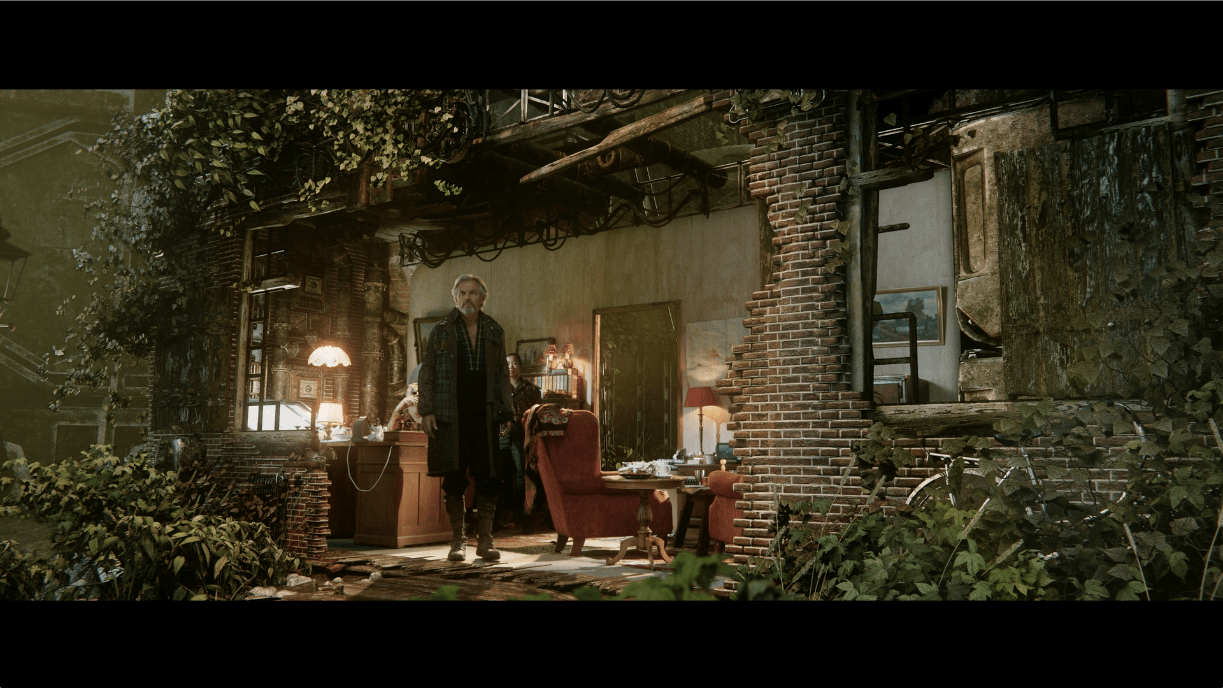}}
\vspace{1mm}
\caption{Example video frames of each of the 15 video contents in LIVE-NFLX-II.}
\label{contents}
\end{figure*}

To design a diverse encoding space, we considered multiple video contents and encoded them at multiple bitrate values (bitrate ladder). We collected 15 video contents, shown in Fig. \ref{contents}, which span a diverse set of content genres, including action, documentary, sports, animation and video games. Notably, the video sequences also contain computer-generated content, such as Blender \cite{blender} animation and video games. The video sources were shot/rendered under different lighting conditions ranging from bright scenes (Skateboarding) to darker scenes (Chimera1102353). There were different types of camera motion, including static (e.g. Asian Fusion and Meridian Conversation) and complex scenes taken with a moving camera, with panning and zooming (e.g. Soccer and Skateboarding). We summarize some of the content characteristics in Table \ref{content_properties}.

\begin{table*}[htp]
\caption{Content characteristics of the video contents in LIVE-NFLX-II.}
\centering
\scalebox{0.85}{
\begin{tabular}{ | >{\centering\arraybackslash}m{3.25cm} | c | >{\centering\arraybackslash}m{9.5cm} |}
\hline
Video Source & ID & Description \\ \hline
AirShow & AS & Camera tracks object of interest, blue sky background. \\ \hline
AsianFusion & AF & Static camera, zoom-in, uniform background. \\ \hline
Chimera1102353 & CD & Static camera on a dark background, medium motion. \\ \hline
Chimera1102347 & CF & Multiple human faces, zoom-in, low motion. \\ \hline
CosmosLaundromat & CL & Blender animation, saliency, low motion, camera panning or static \\ \hline
ElFuenteDance & ED & Rich spatial activity, multiple human faces \\ \hline
ElFuenteMask & EM & Medium spatial activity, saliency \\ \hline
GTA & GTA & Gaming content, fast motion \\ \hline
MeridianConversation & MC & Low-light, human face, low motion, static camera \\ \hline
MeridianDriving & MD & Camera zoom-in, face close-up, low motion, human face \\ \hline
SkateBoarding & SB & Fast motion, complex camera motion, saliency \\ \hline
Soccer & SO & Fast moving camera, rich spatial and temporal activity. \\ \hline
Sparks & SP & Slow camera motion, human face, fire sparks, water \\ \hline
TearsOfSteelRobot & TR & Fast motion, multiple moving objects, complex camera motion \\ \hline
TearsOfSteelStatic & TS & Static camera, human close up, low motion \\ \hline
\end{tabular}
}
\label{content_properties}
\end{table*}

An alternate description of encoding/content diversity is encoding complexity. One approach to describe content is via the spatial and temporal activity (SI-TI) plot \cite{winkler2012analysis}, but we were inclined to use a description that more closely relates to the encoding behavior of each content. Contents with high motion and high spatial activity (textures) tend to be harder to compress, hence subjective scores will generally be lower for those contents, given a fixed number of available bits.

\begin{figure}[tp]
\centerline{
\includegraphics[width=\columnwidth]{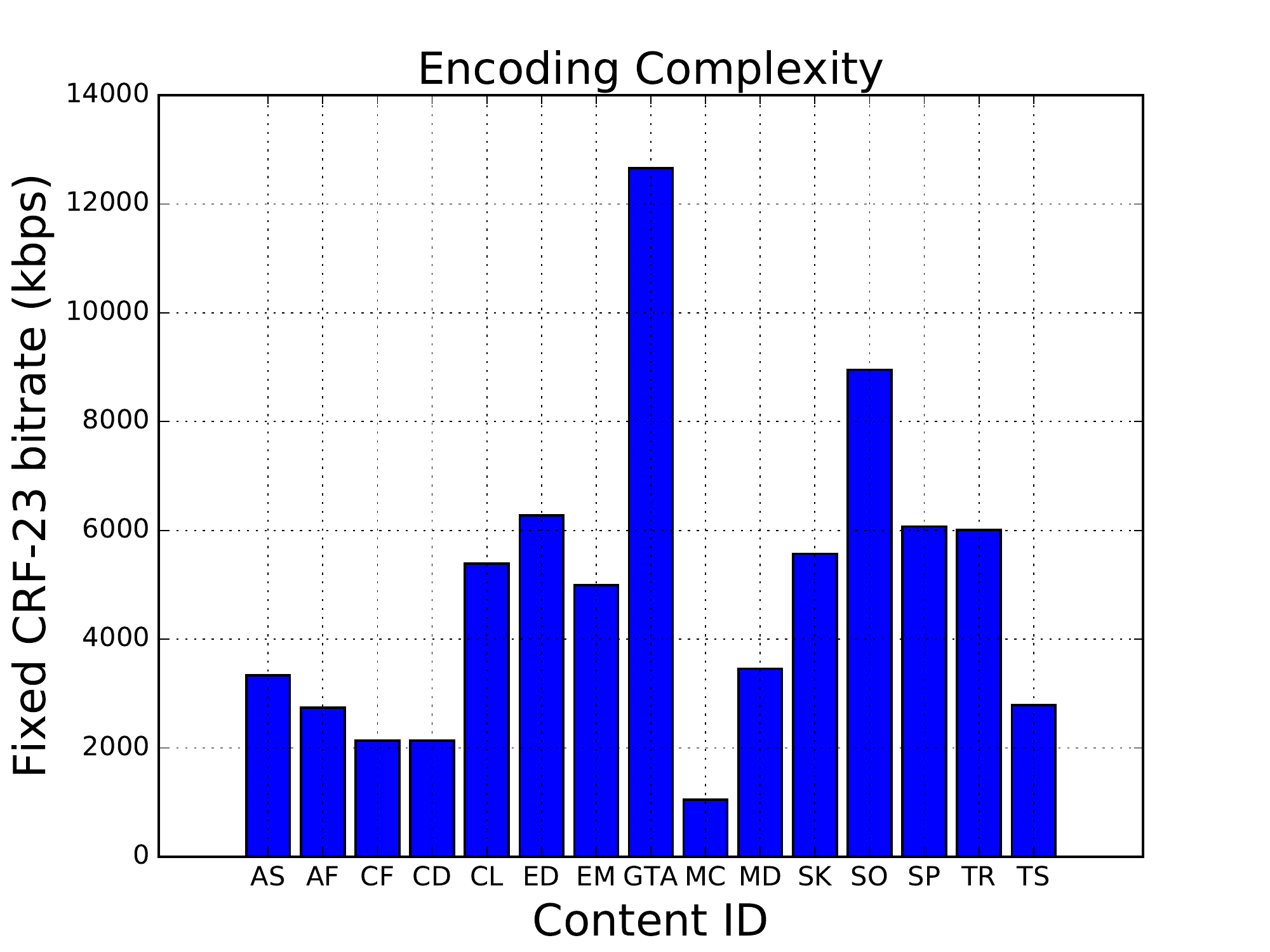}
}
\caption{Content (encoding) complexity for the 15 contents in LIVE-NFLX-II.}
\label{crf_23}
\end{figure}

To measure content encoding complexity, we used the resulting bitrate produced by a constant-quality encoding mode. Specifically, we generated one-pass, fixed constant rate factor (CRF) $=23$ 1920x1080 encodes using libx264 \cite{ffmpeg_x264}, then measured the resulting bitrate (see Fig. \ref{crf_23}). It is clear that there is a large variety of content complexities ranging from low motion contents, such as MeridianConversation or Chimera1102353, medium motion and/or richer textures such as in Skateboarding or ElFuenteMask and high motion and spatial activity as in the Soccer and GTA scenes.

\begin{figure}[tp]
\centerline{
\includegraphics[width=0.85\columnwidth]{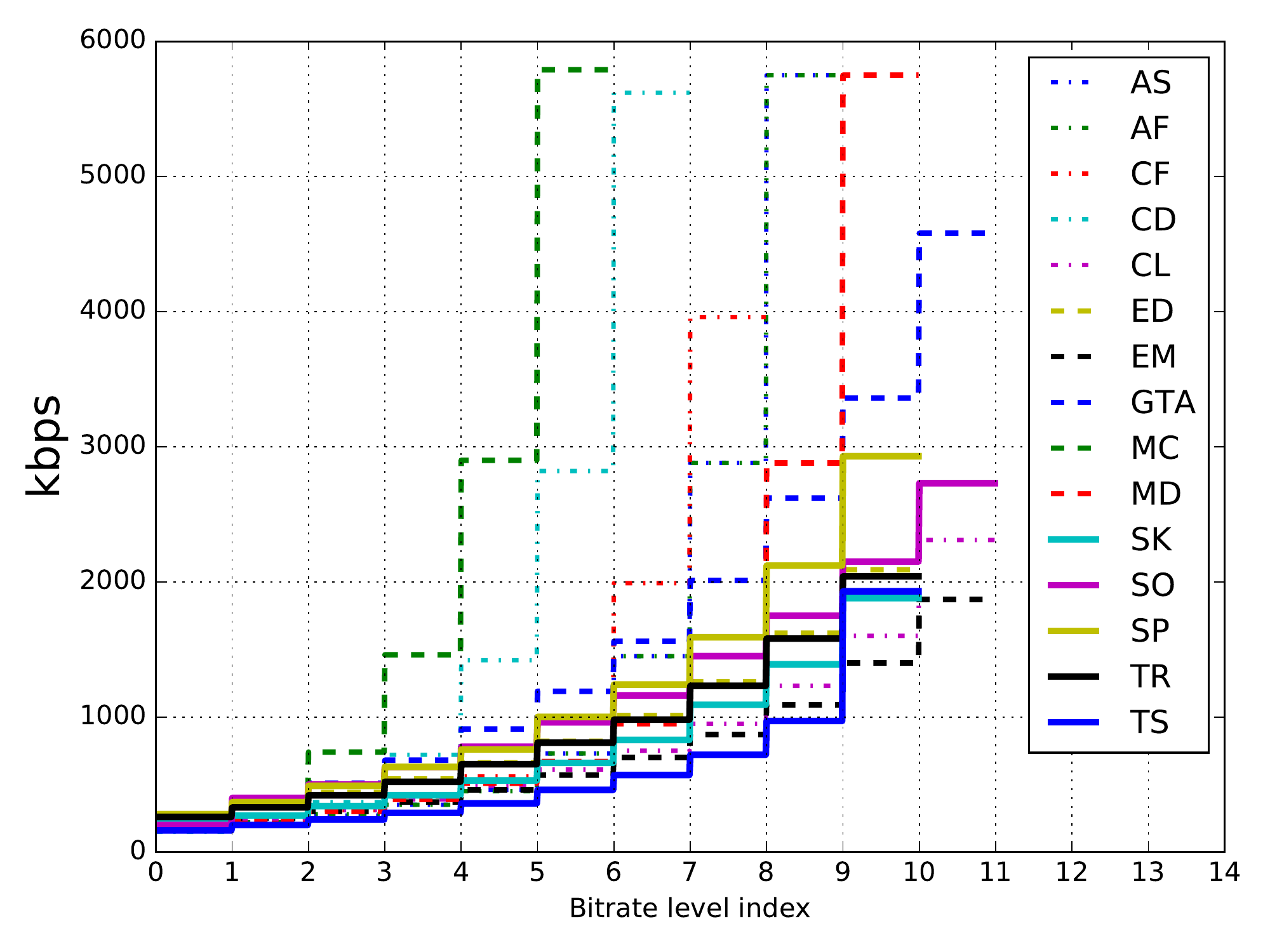}
}
\caption{Bitrate ladder (in Kbps) derived for the LIVE-NFLX-II encodes.}
\label{bitrate_ladder}
\end{figure}

Apart from content complexity, a comprehensive encoding space design also requires a wide range of encoding bitrates and video qualities. To this end, we derived a target bitrate ladder, i.e., a set of possible bitrate values, one for each content. For our content-driven bitrate ladder construction, we used VMAF \cite{techblog} to generate equally spaced (in terms of VMAF) bitrate points, then fed these bitrate points to DO \cite{DO_techblog}. Figure \ref{bitrate_ladder} shows the per-content encoding bitrates that we generated. It may be observed that there is a large variety of encoding rates ranging from about 150 Kbps up to almost 6 Mbps. The low bitrate range, i.e., 150 Kbps to 1 Mbps is sampled more heavily, which aligns well with our raised interest for challenging network conditions.

It should be noted that the video sequences are approximately 25 seconds long and typically contain multiple shots. This design choice is different from commonly used single-shot 10 second test videos, which are widely used in video quality testing. For video streaming applications, we found it more appropriate to use longer video contents with multiple shots, for a number of reasons. Video streaming viewers tend to watch video content that is many minutes long, while the network conditions may vary considerably throughout a streaming session. Having multiple shots also aligns well with the DO encoding approach \cite{DO_techblog}, which leverages the different shot complexities to achieve better encoding efficiency.

\subsection{Network Simulation}

Up to this point, we have only considered the first dimension in the video streaming design space - encoding. Importantly, the number of available bits is not constant in a video streaming session and network resources can vary significantly. To capture the effects of network variability, we manually selected 7 network traces from the HSDPA dataset \cite{riiser2013commute,HSDPA_data}, which contains actual 3G traces collected from multiple travel routes in Norway, using various means of transportation, including car, tram and train, together with different network conditions. This dataset has been widely used to compare adaptation algorithms \cite{mao2017neural} and is suitable for modeling challenging, low-bandwidth network conditions.

\begin{figure}[tp]
\centerline{
\includegraphics[width=0.85\columnwidth]{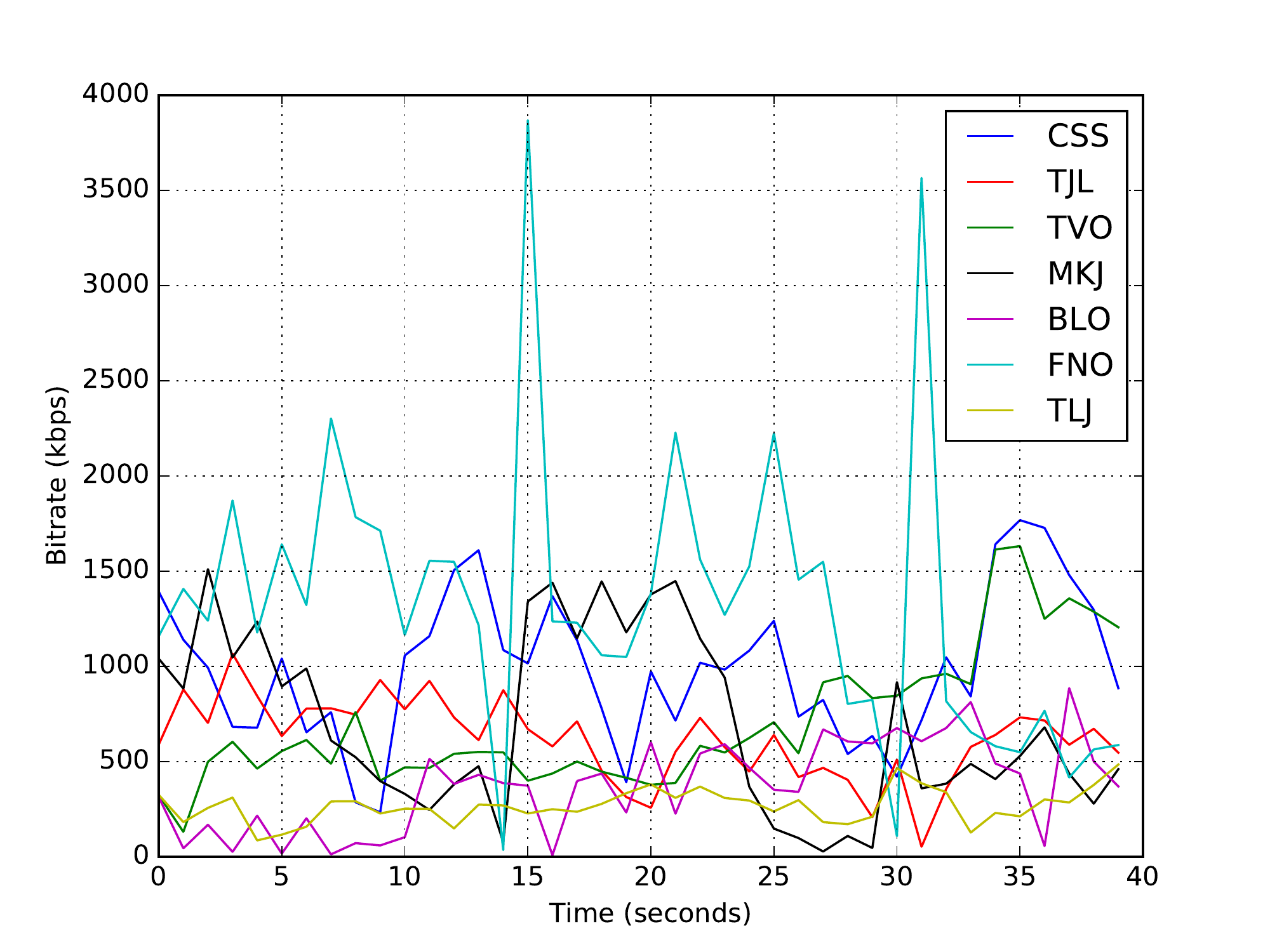}
}
\caption{Real network traces used in our streaming pipeline model.}
\label{network_traces}
\end{figure}

As shown in Fig. \ref{network_traces}, the selected traces are approximately 40 seconds in duration and have varying network behaviors. For example, the TLJ trace has the lowest average bandwidth but does not vary much over time, while the MKJ trace has a much more volatile behavior than TLJ. It may be observed that the network traces densely cover download speeds up to 1Mbps, and there are also samples falling within the 1Mbps-3Mbps range. In Section \ref{network_traces_analytic_table_appendix} of the Appendix, we provide more details about the used network traces.

\begin{figure}[tp]
\centerline{
\includegraphics[width=0.75\columnwidth]{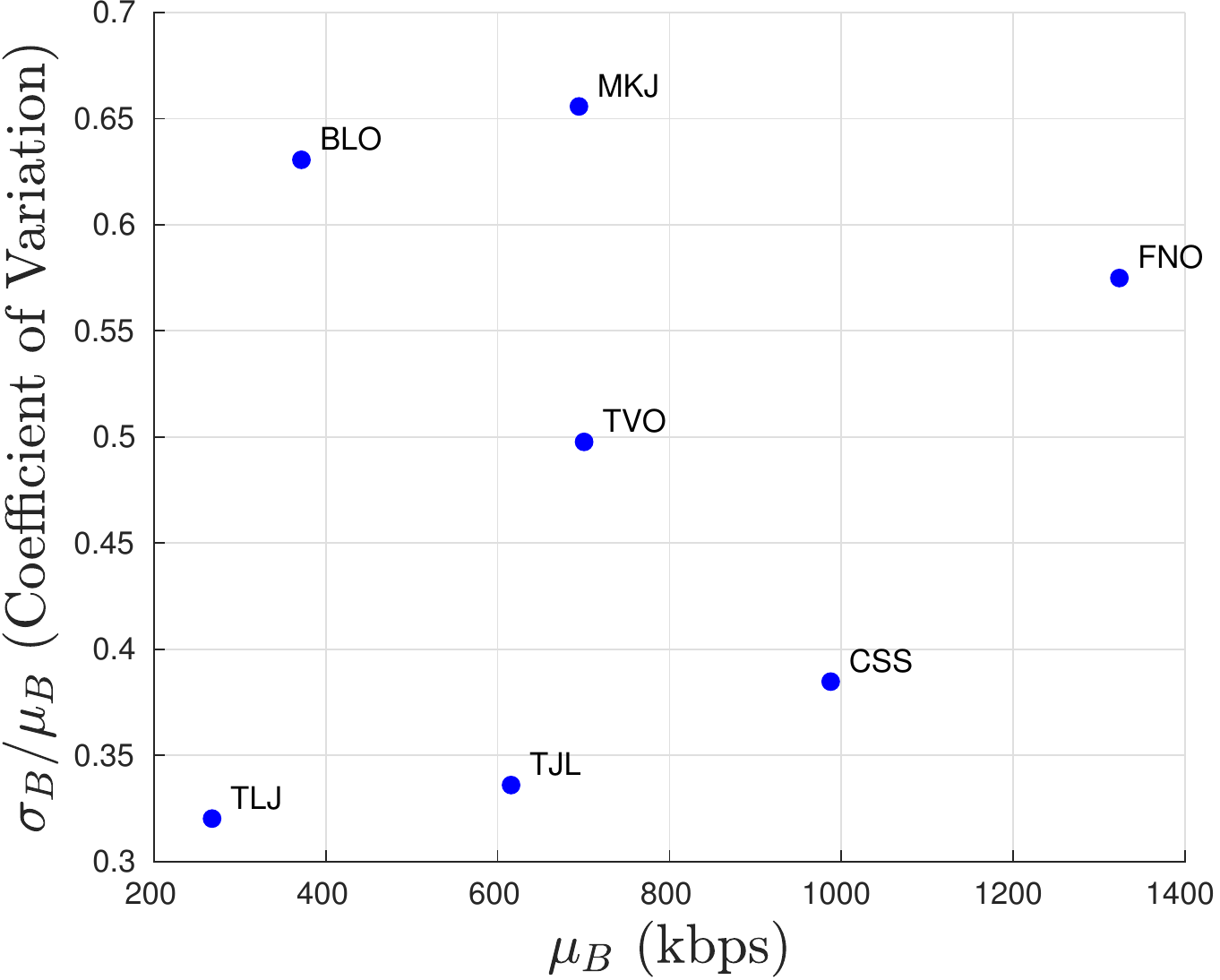}
}
\caption{Plot of $(\mu_B,\sigma_B / \mu_B)$ pairs for the 7 real network traces in LIVE-NFLX-II.}
\label{network_div}
\end{figure}

To model challenging network conditions and train QoE models that can reliably measure QoE under such network conditions, we chose traces that are likely to cause sudden bitrate/quality changes and rebuffers even if the average bandwidth is relatively high. Figure \ref{network_div} shows all 7 pairs of $(\mu_B,\sigma_B/\mu_B)$, where $\mu_B$ and $\sigma_B$ denote the average and standard deviation of the available bandwidth over time. The ratio $\sigma_B/\mu_B$ is also known as the coefficient of variation, which we use to describe network volatility. This design may not necessarily cover all possible combinations (e.g. low $\mu_B$ and high $\sigma_B$ or high $\mu_B$ and low $\sigma_B$), but is challenging in that both low bandwidth conditions and highly varying network conditions are captured. Even for a higher (on average) bandwidth condition, e.g., FNO, the network variations can potentially lead to quality changes and rebuffers.

\subsection{Client ABR Algorithm}

In client-based video streaming, the client is responsible for requesting the next chunk to be played. To decide the appropriate quality representation, the client module is aware of its buffer status, and may estimate future bandwidth (based on past client measurements). The client may also have information regarding the bitrate/quality levels for each video segment. In practice, this can be implemented as part of the manifest exchange between server and client.

Client-based ABR strategies can be broadly classified as: throughput-based \cite{liu2011rate,jiang2014improving,sun2016cs2p}, buffer-based \cite{huang2015buffer,miller2012adaptation,spiteri2016bola,beben2016abma+} and hybrid/control-theoretical approaches \cite{zhou2013buffer,li2014probe,yin2015control,de2013elastic,wang2016squad,mansy2013sabre}. Throughput-based approaches rely on TCP throughput estimates to select subsequent rate chunks, while buffer-based approaches use measurements of buffer occupancy to drive these decisions. Hybrid algorithms use both throughput estimates and buffer occupancy, and deploy control-theoretical or stochastic optimal control formulations to maximize user QoE \cite{yin2015control}. Recently, raw network observations were also fed to neural networks to achieve adaptive rate selection \cite{mao2017neural}. An excellent survey of ABR algorithms is found in \cite{kua2017survey}.

The design space of adaptation algorithms is very large, and hence we selected four representative adaptation algorithms. Each of these four algorithms focuses on different design aspects, such as preserving buffer status, maximizing download bitrate, or mediating between chunk quality and buffer level. Table \ref{acronym_definition} defines some of the acronyms used hereafter.

We implemented the buffer-based (BB) approach from \cite{huang2015buffer}, which decides the rate of the next chunk to be played, as a function of the current buffer occupancy. For BB, a reservoir of $r=5$ sec. and a cushion of $c=4.5$ sec. was used. The advantage of the BB approach is that it can reduce the amount of rebuffering by only accessing buffer occupancy. 

Viewing adaptation from a different perspective, we also implemented a rate-based (RB) approach which selects the maximum possible bitrate such that, based on the estimated throughput, the downloaded chunk will not deplete the buffer. To estimate future throughput, an average of $w=5$ past chunks is computed. Intuitively, selecting $w$ can affect adaptation performance, if the network varies significantly. A low value of $w$ could be insufficient to produce a reliable bandwidth estimate, while a very large $w$ might include redundant past samples and have a diminishing return effect. Another downside of the RB approach is that, when channel bandwidth varies significantly, it may lead to excessive rebuffering and aggressive bitrate/quality switching.

Using video bitrate as a proxy for quality may yield sub-optimal results; a complex shot (rich in spatial textures or motion) requires more bits to be encoded at the same quality compared to a static shot having a uniform background and low motion. Therefore, it is interesting to explore how well a quality-based (QB) adaptation algorithm will correlate against subjective scores. We relied on the dynamic programming consistent-quality adaptation algorithm presented in \cite{li2014streaming}. We use VMAF measurements (using the video quality module - see Section \ref{video_quality_module_appendix}) as a utility function to be maximized within a finite horizon $h$ (in sec.). This utility maximization is formulated as a dynamic programming (DP) problem solved at each step, which determines the chunk to be played next.

In our QB implementation, the network conditions are estimated similar to our RB implementation. We assume that future throughput (within the horizon $h$) will be equal to the average throughput over the past $w=5$ chunks. However, different from RB, QB maximizes visual quality in terms of VMAF, instead of video bitrate. For the QB client, two practical limitations on the buffer size are imposed. To reduce the risk of rebuffering, the QB solution requires that the buffer is never drained below a lower bound $B_l$ (in sec.). Also, due to physical memory limitations, QB never fills the buffer above a threshold $B_h$. To ensure that the $B_l$ and $B_h$ constraints are satisfied, the QB solution is set to converge to a target buffer $B_t \in (B_l, B_h)$ by imposing in its DP formulation that the buffer at the end of the time horizon has to be equal to $B_t$. Notably, if the dynamic programming solution fails (when $B_l$ cannot be achieved or $B_h$ is surpassed), the QB algorithm uses a ``fallback" mode. In particular, if $B_l$ cannot be achieved, then QB selects the lowest quality stream, while if $B_h$ is surpassed, then QB pauses downloading until the buffer frees up and then downloads the highest available stream.

It is impossible for any adaptation strategy to have perfect knowledge of future network conditions. In practice, probabilistic network modeling, or other much simpler estimation techniques can be exploited. For the latter, many adaptation algorithms assume that network conditions are constant over short time scales, and apply filtering using previous network measurements, as in QB. Since accurate knowledge of future bandwidth places an upper bound on the performance of an algorithm, we also included a version of QB which uses the actual network traces, instead of throughput estimates, thereby acting as an ``oracle" (OQB).

\begin{table*}[htp]
\caption{Acronym definition table.}
\centering
\scalebox{0.85}{
\begin{tabular}{| c | c | c | c | c |}
\hline
Acronym & Definition & Measured in & Value & Used in  \\ \hline
BB & buffer-based adaptor & - & - & - \\ \hline
RB & rate-based adaptor & - & - & - \\ \hline
QB & quality-based adaptor & - & - & - \\ \hline
OQB & oracle quality-based adaptor & - & - & - \\ \hline
$B_0$ & pre-fetched video data & \# chunks & 1 & BB, RB, QB, OQB \\ \hline
$B_l$ & min allowed buffer size & sec. & 1 & QB, OQB \\ \hline
$B_h$ & max allowed buffer size & sec. & 10 & QB, OQB \\ \hline
$T_a$ & actual throughput & Kbps & varies & BB, RB, QB, OQB \\ \hline
$h$ & horizon & sec. & 10 & QB, OQB \\ \hline
$B_t$ & target buffer & sec. & 3 & QB, OQB \\ \hline
$r$ & reservoir for BB & sec. & 5 & BB \\ \hline
$c$ & cushion for BB & sec. & 4.5 & BB \\ \hline
$w$ & window for throughput estimation & \# chunks & 5 & RB, QB, OQB \\ \hline
\end{tabular}}
\label{acronym_definition}
\end{table*}

\begin{figure}[tp]
\centerline{
\includegraphics[width=0.75\columnwidth]{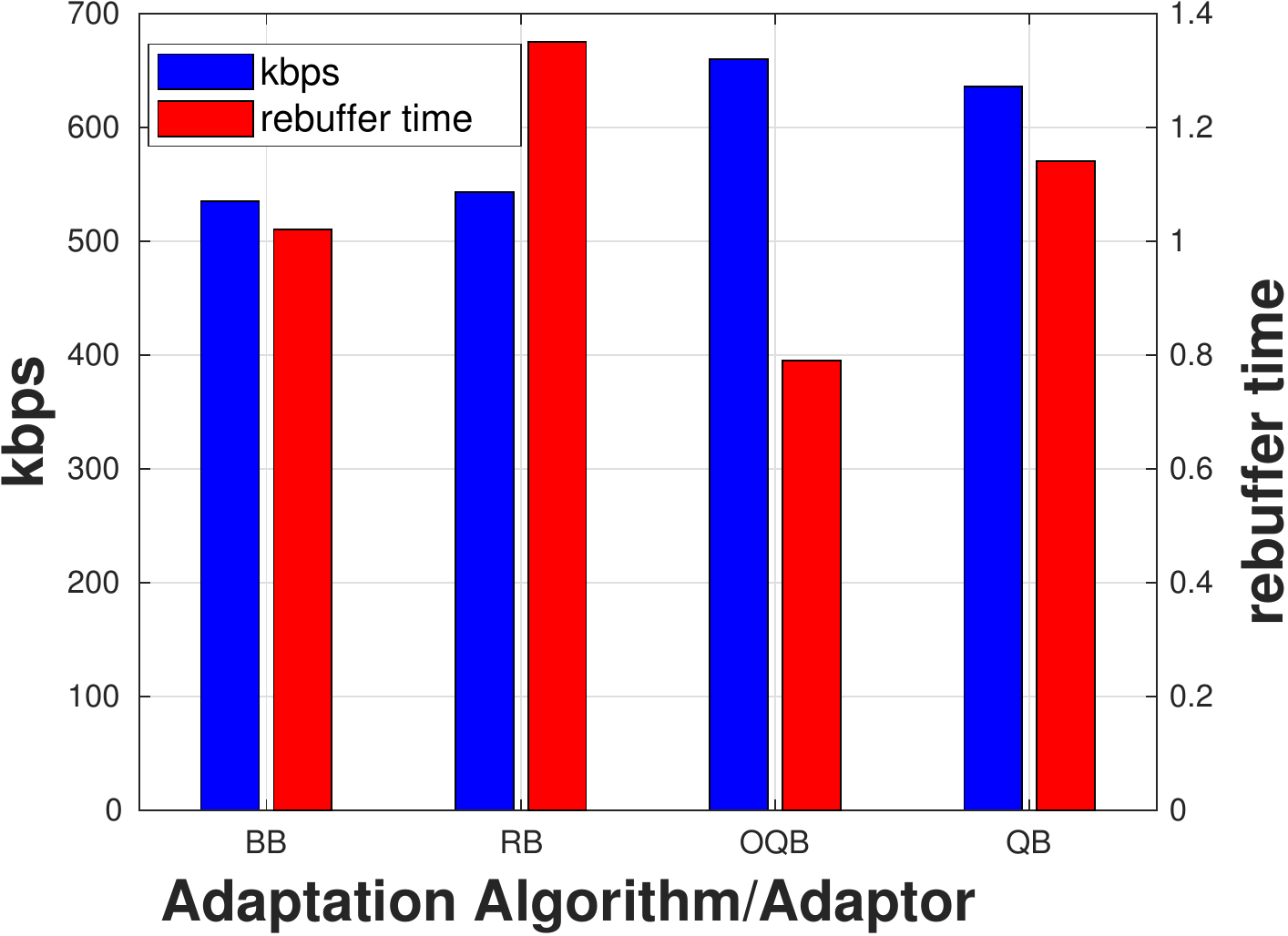}
}
\caption{Average bitrate (in Kbps) and rebuffering time for the 4 adaptation algorithms in LIVE-NFLX-II.}
\label{ABR_div}
\end{figure}

To demonstrate the diversity in ABR algorithms, Fig. \ref{ABR_div} shows the average bitrate (in Kbps) and rebuffering time for the 4 adaptation algorithms. We observed bitrate values in the range of 535 to 660 Kbps and average rebuffering times from 0.8 to 1.35 seconds (see also Table \ref{Objective_Aggregate_Analysis}). We revisit the ABR algorithms by studying more QoE indicators in Section \ref{ABR_objective_analysis}.

Given the comprehensive nature of the encoding, network conditions and ABR designs, we are able to create a rich streaming QoE database by conducting a large subjective test on human perception. In the next Section, we describe the specifics of this test, which led to the creation of the LIVE-NFLX-II database.

\section{Subjective Test on the Recreated Experience}
\label{subjective_test}

A single-stimulus continuous quality evaluation study \cite{BT50013} was carried out over a period of four weeks at The University of Texas at Austin's LIVE subjective testing lab. We collected retrospective and continuous-time QoE scores on a 1080p 16:9 computer monitor from a total of 65 subjects. To collect continuous scores, a rating bar was displayed on the bottom of the screen throughout the video playout, allowing subjects to report their QoE in real time. Retrospective scores reflect the overall QoE after viewing each video sequence in its entirety, while continuous scores capture the time-varying nature of QoE due to quality changes and rebuffering. The subjects were comfortably seated two feet away from the computer monitor. This viewing distance corresponds to approximately three times the height of the computer monitor.

Given the large number of videos (420 in total) to be evaluated and necessary constraints on the duration of a subjective study, we showed only a portion of the distorted videos to each subject via a round-robin approach as follows. Each subject viewed all 15 contents, but only 10 distorted (2 adaptors and 5 network traces) videos per content. We assumed the sequence of adaptors BB, RB, QB and OQB and network traces 0 to 7, then assigned them to subjects in a circular fashion. For example, if subject $i$ was assigned to adaptors BB and RB and network traces 0 to 4, then subject $i+1$ was assigned to adaptors RB and QB and traces 1 to 5. This led to a slightly uneven distribution of subjects per distorted video, but we considered this to have a minor effect. The benefit of a round robin approach compared to a random assignment is that we can have guaranteed coverage for all traces and adaptors.

To avoid user fatigue, the study was divided into three separate 30-minute viewing sessions of 50 videos each (150 videos in total per subject). Each session was conducted at least 24 hours apart to minimize subject fatigue \cite{BT50013}. To minimize memory effects, we ensured that within each group of 7 displayed videos, each content was not displayed more than once. We used the Snellen visual acuity test and ensured that all participants had normal or corrected-to-normal vision. Overall, the final database consists of 420 distorted videos (15 contents, 7 network traces and 4 adaptation algorithms) and an average of 23.2 scores (retrospective and continuous) for every distorted video. No video was viewed by less than 22 subjects, ensuring a sufficient number of scores per video. Overall, we gathered 65*150 = 9750 retrospective scores and 9750 continuous-time waveforms to study subjective QoE. It should be noted that the number of subjects and distorted videos in the database is significantly larger than many existing subjective databases.

To design the experimental interface, we relied on Psychopy, a Python-based software \cite{peirce2007psychopy}. Psychopy makes it possible to generate and display visual stimuli with high precision, which is very important when collecting continuous, per-frame subjective data. To facilitate video quality research, we make our subjective experiment interface publicly available at \href{https://github.com/christosbampis/Psychopy_Software_Demo_LIVE_NFLX_II}{https://github.com/christosbampis/Psychopy\_Software\_Demo\_\\LIVE\_NFLX\_II}. Using the interface, we collected opinion scores (retrospective and continuous) in the range of [1, 100].

Following data collection, we applied z-score normalization per subject and per session \cite{BT50013} to account for subjective differences when using the rating scale. To reliably calculate the retrospective Mean Opinion Score (MOS), we applied subject rejection on the z-scored values according to \cite{BT50013}. After subject rejection, we found that the retrospective scores were in high agreement, exhibiting a between-group (splitting the scores per video into two groups and correlating) Spearman's Rank Order Correlation Coefficient of 0.96. For the continuous scores, we performed a similar z-score normalization and then averaged the normalized continuous scores per subject to compute a continuous MOS score for each frame. While more advanced subject rejection techniques could have been used as in \cite{bampis2017study}, we found that the average (across subjects) continuous-time scores did not significantly change after continuous-time rejection.

\section{Objective Analysis of LIVE-NFLX-II}
\label{objective_analysis}

Before studying the human subjective data in the database, it is interesting to analyze the generated video streams using simple QoE indicators, like video quality or buffer level for various dimensions of the design space. First, we will present how video quality, measured in terms of VMAF, is affected on each content and demonstrate the content/encoding diversity in the database (Section \ref{objective_content_analysis}). Then, we analyze the network traces and the adaptation algorithms that were tested (Sections \ref{objective_network_analysis} and \ref{ABR_objective_analysis} respectively).

\subsection{Content Analysis}
\label{objective_content_analysis}

\begin{figure}[tp]
\centerline{
\includegraphics[width=0.85\columnwidth]{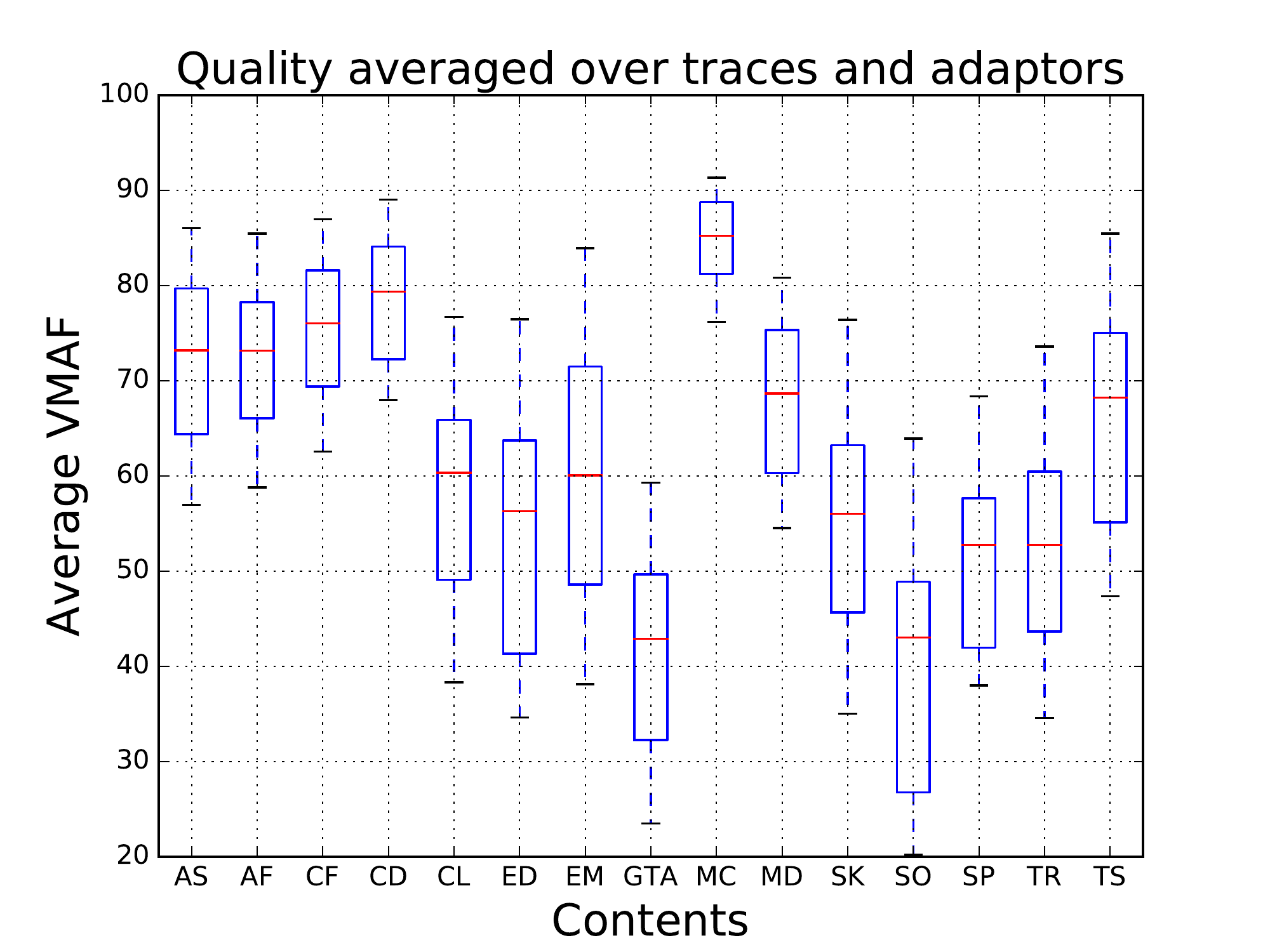}
}
\caption{Averaged (over segments, traces and adaptors) VMAF values of the 15 contents in LIVE-NFLX-II. The rebuffering intervals were not taken into consideration when making this plot. The box extends from the lower to upper quartile values of the data, with a line at the median. The whiskers extend from the box to show the range of the data \cite{python_boxplot_web}. The same boxplot notation is used in Fig. \ref{FinalMOS_Adaptor_Analysis}.}
\label{vmaf_per_content}
\end{figure}

Besides constructing a bitrate ladder, which is typically carried out on the server side, we can also measure the end-user quality received on the client device. Given that bitrate is not sufficient to capture perceptual quality, we use the VMAF perceptual index \cite{techblog}. We used VMAF to measure video quality over all 420 videos and averaged the values for each content, as shown in Fig. \ref{vmaf_per_content}. It can be observed that contents having low complexities, such as MC, CF and CD, were delivered with better VMAF values. By contrast, challenging content, like GTA and Soccer (SO), were streamed at significantly lower quality. This reveals the importance of content-driven encoding on the server and the potential of content-aware streaming strategies, where encoding/streaming parameters are customized to the video content streamed by each client. We revisit some of the details behind our encode generation (encoding module) in Section \ref{encoding_module_appendix}.

\subsection{Network Condition Analysis}
\label{objective_network_analysis}

\begin{figure}[tp]
\centerline{
\includegraphics[width=\columnwidth]{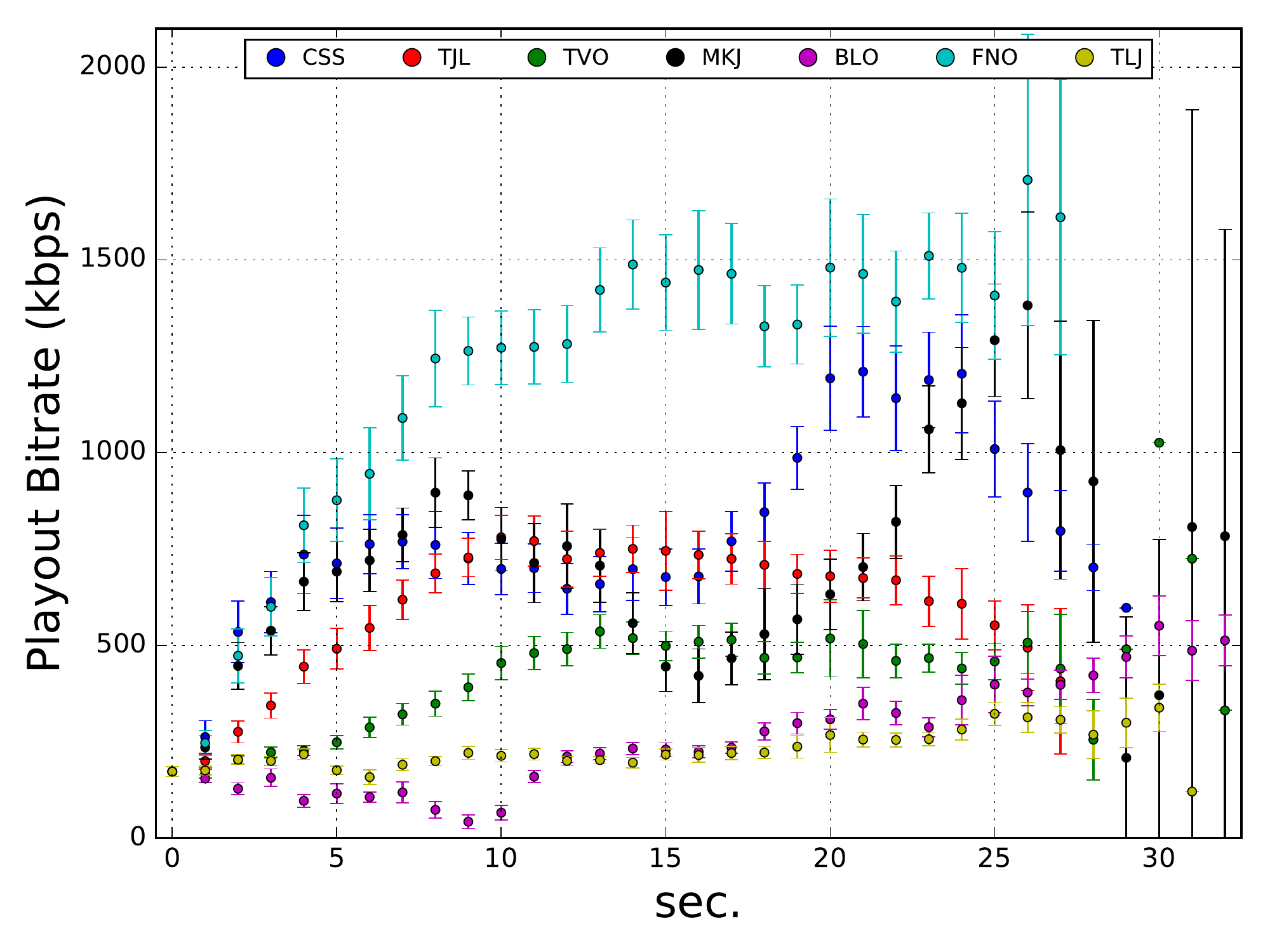}
}
\caption{Playout bitrate over time across different network traces. To capture the effects of rebuffering intervals, a value of 0 is used for the video bitrate during those time instants. The error bars indicate the $95\%$ confidence interval.}
\label{bitrate_per_trace}
\end{figure}

To analyze the behavior across different network traces, we first collected measurements of the playout bitrate, averaged it over each second (and across contents and adaptors) and present its dynamic per network condition evolution (average and 95$\%$ confidence intervals) in Fig. \ref{bitrate_per_trace}. Note that since video contents are at most 25 seconds long, only sessions that experienced rebuffering lasted longer than 25 seconds. Thus fewer samples are available after 25 seconds and the confidence intervals become larger.

As expected, better network conditions overall (FNO and CSS) achieved better playout bitrates when compared to low-bandwidth cases, as in TLJ and BLO. It is interesting to observe that volatile traces, such as MKJ and CSS, led to significant differences in bitrate, but this was not the case for FNO. Since FNO provides better network throughput on average than MKJ and CSS, the video buffer was sufficiently filled to account for sudden drops.

\subsection{Adaptation Algorithm Analysis}
\label{ABR_objective_analysis}

To study adaptation behavior, we first collected key metrics (e.g. number of rebuffers) for all distorted videos generated by each adaptation algorithm. Table \ref{Objective_Aggregate_Analysis} shows that the OQB adaptor improved most of the objective streaming metrics, e.g. the playout bitrate or the rebuffering duration. This demonstrates that a better bandwdith prediction model can improve the behavior of an ABR algorithm.

By contrast, the RB adaptor led to the largest amount of rebuffering, since it picks the subsequent chunk rate in a greedy fashion, it is myopic (does not look ahead in time) and does not take into account the buffer status. The more conservative BB reduces the amount of rebuffering as compared to RB and QB, and has the least number of quality switches. Nevertheless, given that it does not explicitly seek to maximize bitrate, it delivers the lowest playout bitrate. Between RB and BB, QB offers a better tradeoff between playout bitrate and rebuffering. These results are not very surprising: maximizing quality/bitrate or avoiding rebuffering are conflicting goals, and hence, designing adaptation algorithms should focus more on jointly capturing these factors, as in the case of QB.

At this point, let us take a step back and consider why OQB, despite knowing the entire trace, also suffers from rebuffering. In fact, by setting the maximum buffer $B_h=10$ sec., and $h=10$ sec. the dynamic programming solution may fail to return an optimal solution. We found that by increasing $B_h$ and $h$, both OQB and QB could reduce rebuffering, but we purposely decided to challenge the behavior of these ABR algorithms by selecting low/volatile bandwidth traces and a low buffer size. By doing so, we were able to collect valuable subjective responses on video sequences under difficult streaming conditions, including significant video quality degradations and rebuffering.


\begin{table}[htp]
\caption{Objective comparison between adaptation algorithms. Each attribute is averaged over all 105 videos (15 contents and 7 traces) per adaptor. The bitrate values are imputed with a value of 0 during rebuffering intervals, while the VMAF values are calculated only on playback frames. We use boldface to denote the best adaptation algorithm in each case.}
\centering
\scalebox{1}{
\begin{tabular}{| c | c | c | c | c |}
\hline
Description & BB & RB & OQB & QB \\ \hline
\# switches & \textbf{5.91} & 7.08 & 8.13 & 8.45 \\ \hline
bitrate (Kbps) & 535 & 543 & \textbf{660} & 636 \\ \hline
\# rebuffers & 0.75 & 1.57 & \textbf{0.70} & 0.99 \\ \hline
rebuffer time (sec.) & 1.02 & 1.35 & \textbf{0.79} & 1.14 \\ \hline
per chunk avg. VMAF & 58.05 & 62.58 & \textbf{64.52} & 63.19 \\ \hline
per chunk VMAF diff. & 9.67 & 7.51 & \textbf{6.89} & 8.59 \\ \hline
\end{tabular}}
\label{Objective_Aggregate_Analysis}
\end{table}

\begin{figure*}[tp]
\centering
\subcaptionbox{Playout Bitrate}{\includegraphics[width=\columnwidth]{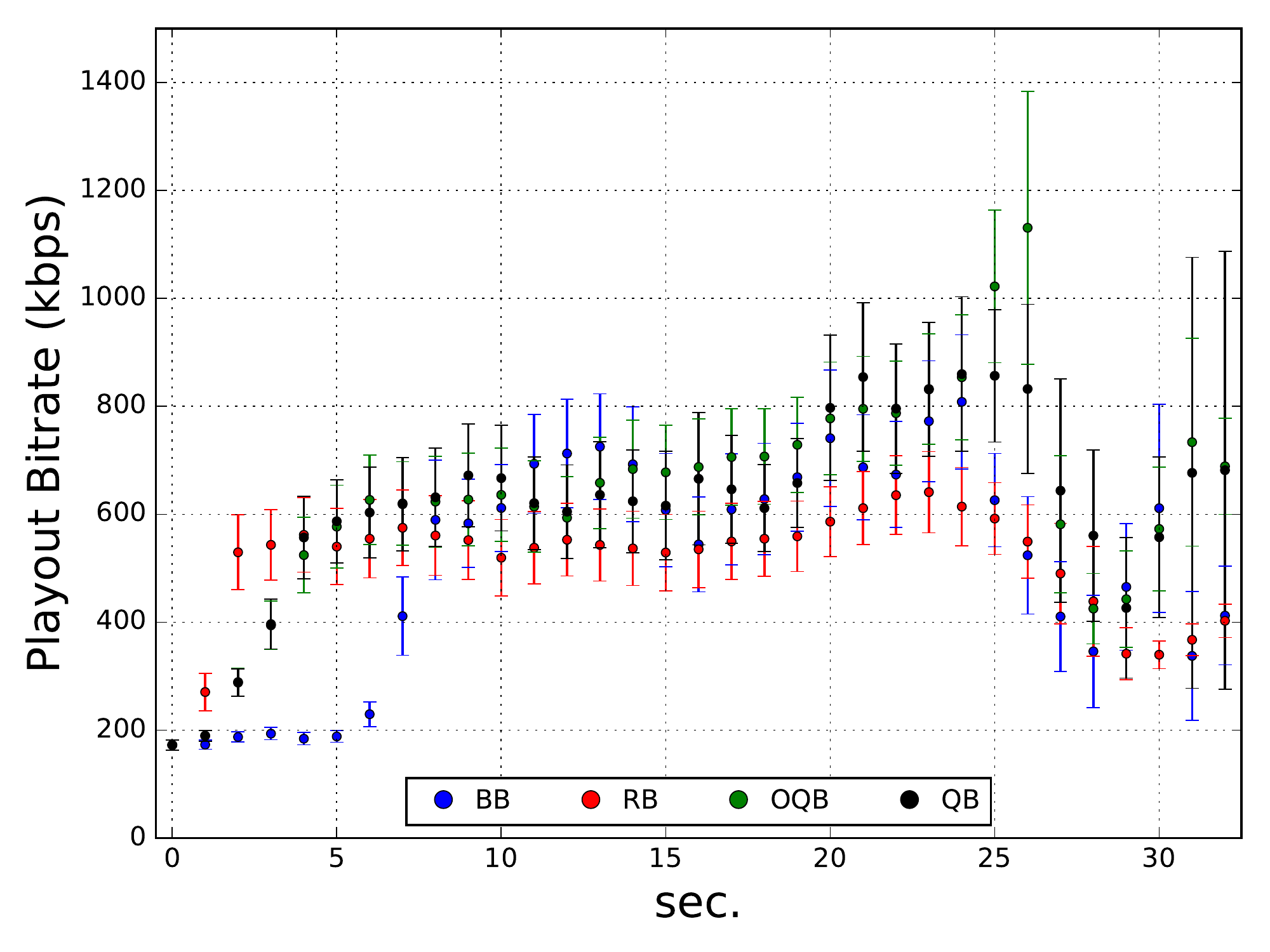}}\hfill
\subcaptionbox{Buffer Level}{\includegraphics[width=\columnwidth]{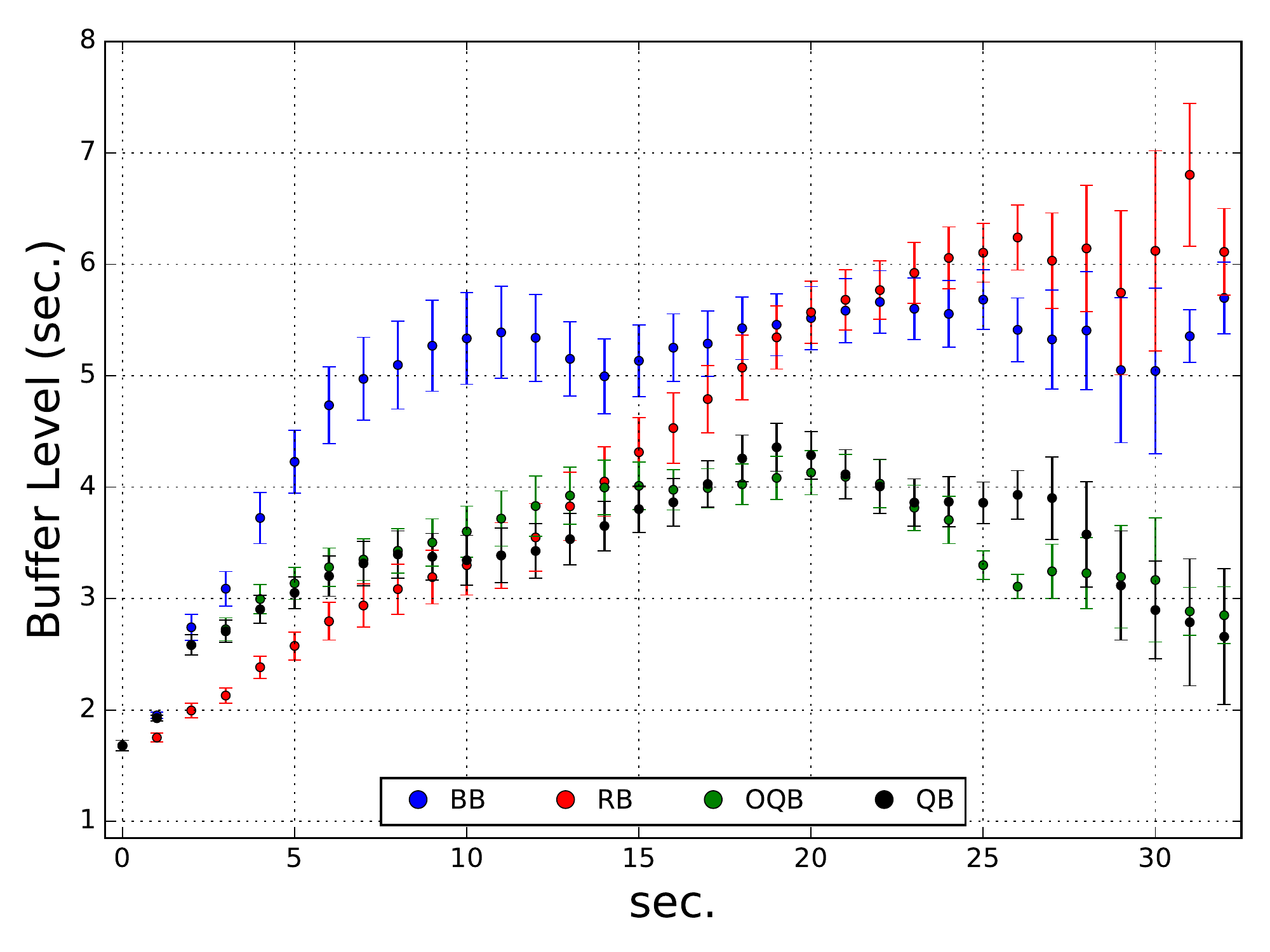}}
\caption{Playout bitrate and buffer level over time for different adaptation algorithms (averaged across traces and contents). To capture the effects of rebuffering intervals, a value of 0 is used for the video bitrate during those time instants. The error bars indicate the $95\%$ confidence interval.}
\label{ContinuousTime_Objective_Analysis}
\end{figure*}

Let us now further dig into how each adaptation algorithm behaves over time within each session. As before, we measure the per second playout bitrate and buffer level, and show the per adaptation evolution in Fig. \ref{ContinuousTime_Objective_Analysis}. In terms of bitrate, RB starts aggressively for the first few seconds, but then tends to have a lower bitrate compared to quality-based adaptors. By contrast, BB is the most conservative strategy in terms of bitrate, while QB and OQB deliver start-up bitrates in-between RB and BB. However, after about 15 seconds, QB and OQB consistently deliver higher bitrates. Note that video can only be longer than 25 seconds due to rebuffers. Therefore, for time intervals greater than 25 seconds, we are dealing with video sequences under more challenging conditions and hence it is expected that the bitrate decreases and the buffer level decreases or stays the same.

More importantly, Fig. \ref{ContinuousTime_Objective_Analysis} shows that ABR algorithms mainly differ from each other during the start-up phase. For example, RB chooses playout bitrate aggressively by closely following the available bandwidth, while BB is more conservative and prioritizes on accumulating video buffer. After the start-up phase, the ABR algorithms converge and make similar decisions.

\begin{figure}[htp]
\centerline{
\includegraphics[width=\columnwidth]{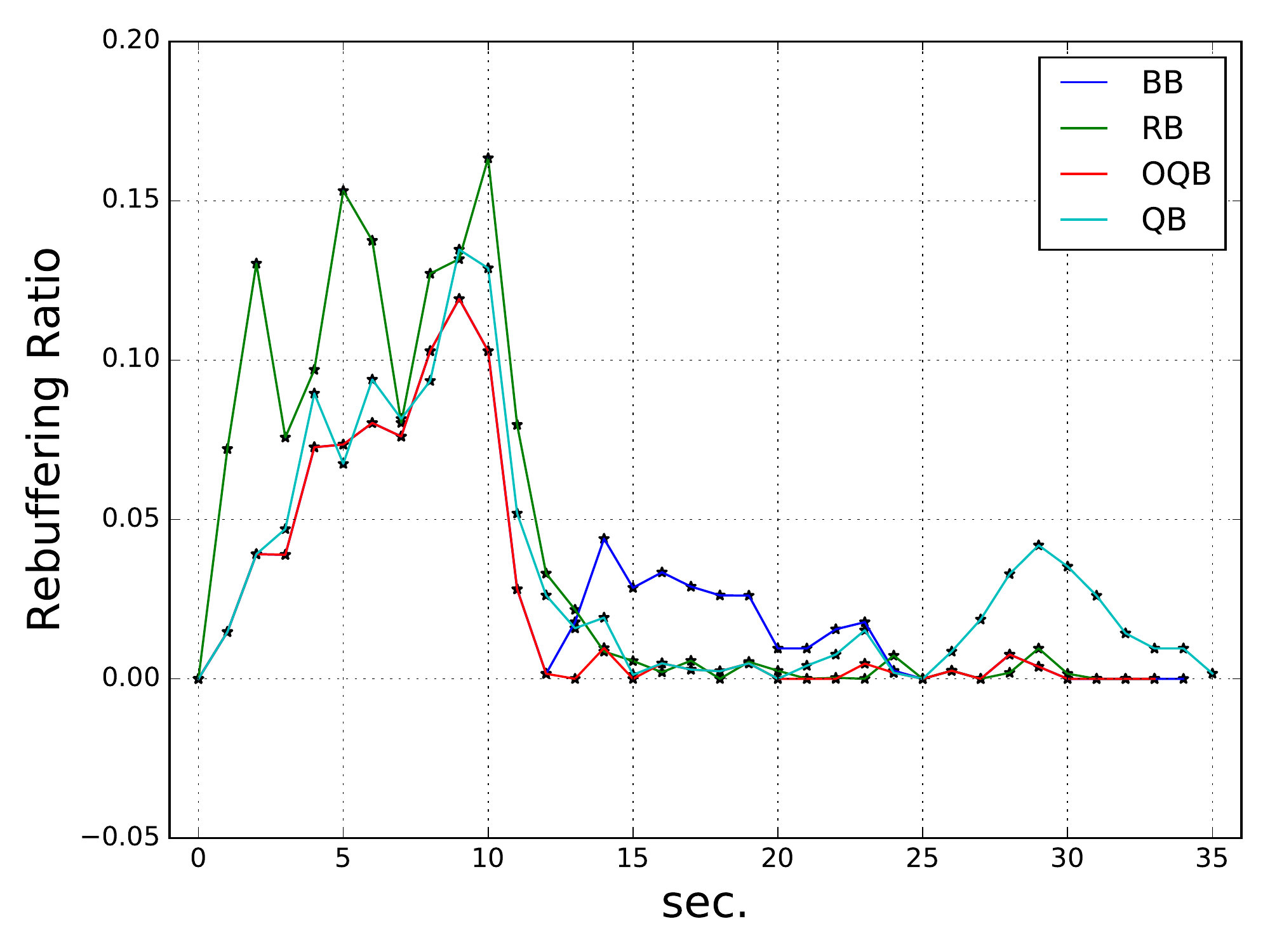}
}
\caption{Rebuffering time instants for different adaptation algorithms (averaged across traces and contents). The rebuffering ratio (vertical axis) is the fraction of videos (out of 105 videos per adaptation algorithm) that rebuffered within a one second window.}
\label{RebufferingLocation_Analysis}
\end{figure}

During the start-up phase, there is little to no video buffer to absorb the impact of network variations and hence different ABR decisions will also lead to different buffer levels. For example, the combination of aggressive quality switching and network volatility leads RB to produce the worst rebuffering in the start-up phase (see also Fig. \ref{RebufferingLocation_Analysis}) and slows down buffer build-up (Fig. \ref{ContinuousTime_Objective_Analysis}b). Nevertheless, after sufficient time, the RB buffer level increases and even surpasses the BB buffer level. In the case of QB and OQB, both adaptors try to reach the target buffer $B_t=3$. We note that for the RB and BB adaptors we did not specify a maximum buffer size, but as shown in Fig. \ref{ContinuousTime_Objective_Analysis}b, none of the ABR algorithms achieve $B_h=10$ sec., given the challenging network traces.

Until now, we have been mostly comparing ABR algorithms. Nevertheless, we have also found a very important similarity: rebuffering events tend to occur earlier in the video playout. To demonstrate this, we calculated  the rebuffering ratio of each adaptor over time, i.e., the average rebuffering rate incurred by an adaptor throughout the playout. Figure \ref{RebufferingLocation_Analysis} shows that all adaptors have significantly higher rebuffering ratios early on, since the buffer is not yet filled. This is also related to the fact that we only fetch one chunk before starting the playout (see Appendix, Section C). Between adaptors, there are, of course, differences as well. The RB adaptor experiences heavier early rebuffering, since the buffer level is not taken into account. Between QB and OQB, the difference is that QB can lead to rebuffering much later in the video, while OQB, which is aware of the entire network trace, is able to minimize rebuffering events from occurring at a later time.

\section{Subjective Analysis}
\label{subjective_analysis}

\begin{figure*}
\centering
\subcaptionbox{VMAF and MOS}{\includegraphics[width=0.66\columnwidth]{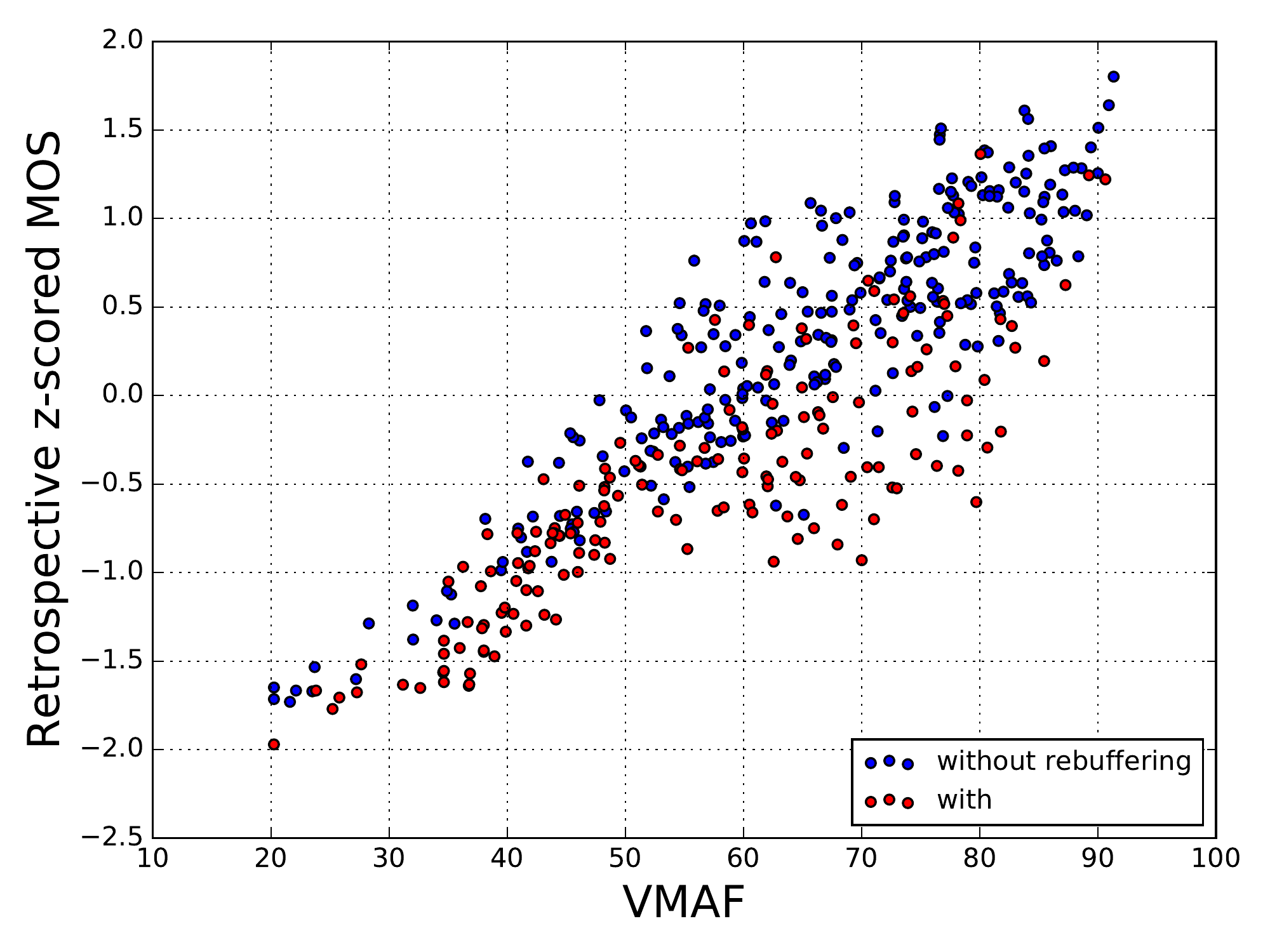}}\hfill
\subcaptionbox{\# rebuffers and MOS (95\% conf. intervals)}{\includegraphics[width=0.66\columnwidth]{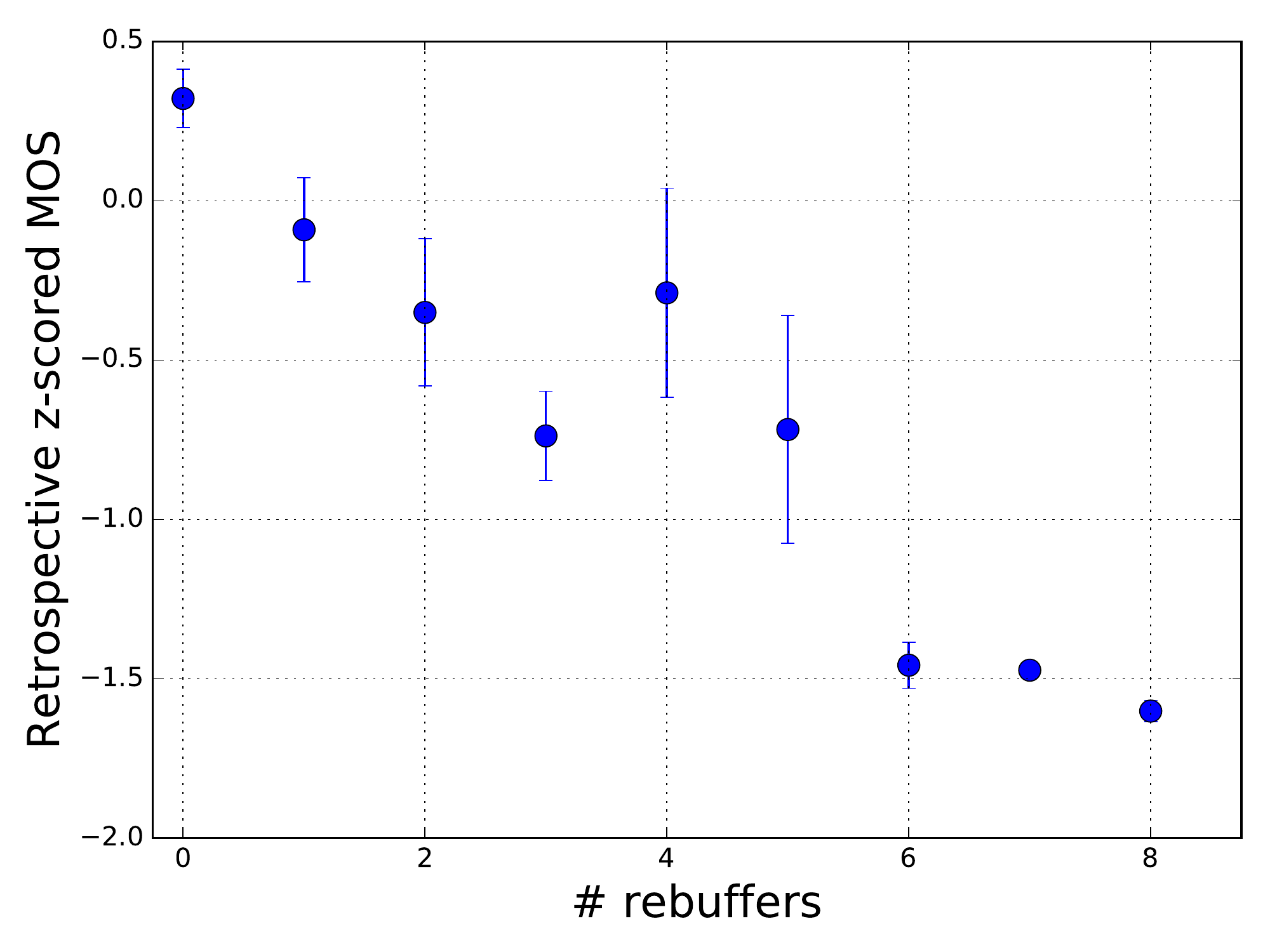}}\hfill
\subcaptionbox{Rebuffer duration and MOS}{\includegraphics[width=0.66\columnwidth]{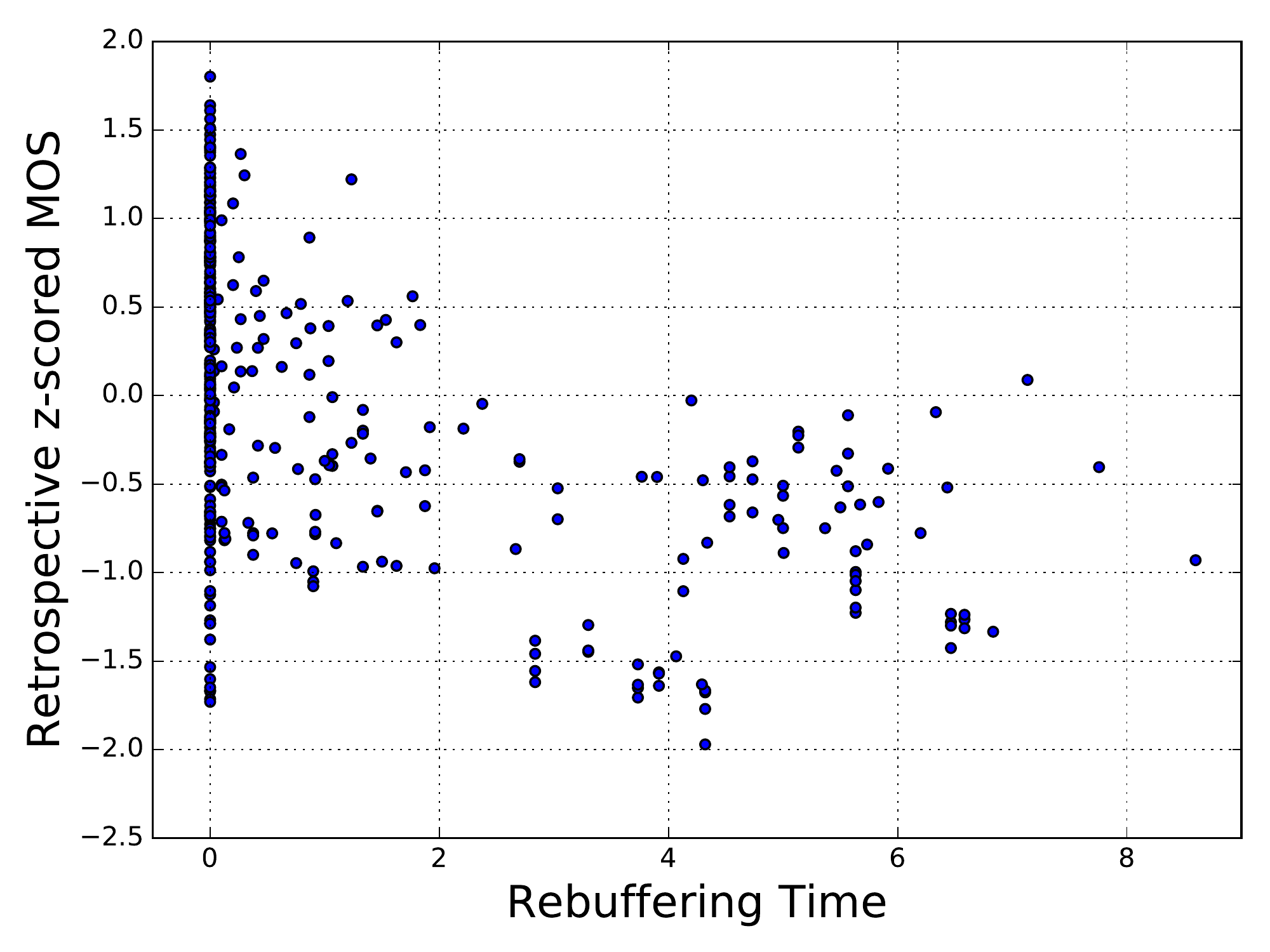}}
\caption{VMAF measurements, number and duration of rebuffer events against retrospective opinion scores in LIVE-NFLX-II. Around $40\%$ of the video sequences have at least one rebuffering event.}
\label{FinalMOS_Analysis}
\end{figure*}

Up to this point, we have studied the behavior of different network traces and adaptors with respect to some QoE-related factors. Nevertheless, in streaming applications, human opinion scores serve as the ground truth when analyzing streaming video impairments and when evaluating objective models of video quality and QoE prediction. Here we analyze the video database by means of the collected retrospective and continuous-time subjective scores.

\subsection{Analysis Using Retrospective Scores}

To identify the main QoE factors, Fig. \ref{FinalMOS_Analysis} highlights the relationships between retrospective scores and average VMAF values (calculated on non-rebuffered frames), and the number and duration of rebuffering events respectively. Unsurprisingly, the presence of rebuffering (red points) negatively impacts the overall correlation of VMAF with subjective opinion scores, since VMAF does not account for the effects of rebuffering on user experience. In Section \ref{objective_VQA_QoE}, we show how QoE prediction models based on VMAF can deliver improved performance. Meanwhile, it can be seen that a larger number of rebuffering events tends to decrease user experience. 

As an exception, the points with 3, 4 and 5 rebuffering events are not in decreasing MOS order. Interestingly, for these points, we found that the average rebuffering duration was 4.33, 3.49 and 2.93 sec. respectively, meaning that larger rebuffering occurrence did not necessarily imply larger rebuffering duration. Therefore, Fig. \ref{FinalMOS_Analysis}b demonstrates that subjects are sensitive to a combined effect of rebuffering occurrence and duration.

In Fig. \ref{FinalMOS_Analysis}c, we observe that a longer rebuffering time also lowers QoE, but when the rebuffering time is more than 4 seconds, \textit{duration neglect effects} \cite{hands2001recency} may reduce this effect. According to the duration neglect phenomenon, subjects may recall the duration of an impairment, but they tend to be insensitive to its duration (after a certain cutoff) when making retrospective QoE evaluations.

\begin{figure}[htp]
\centerline{
\includegraphics[width=0.95\columnwidth]{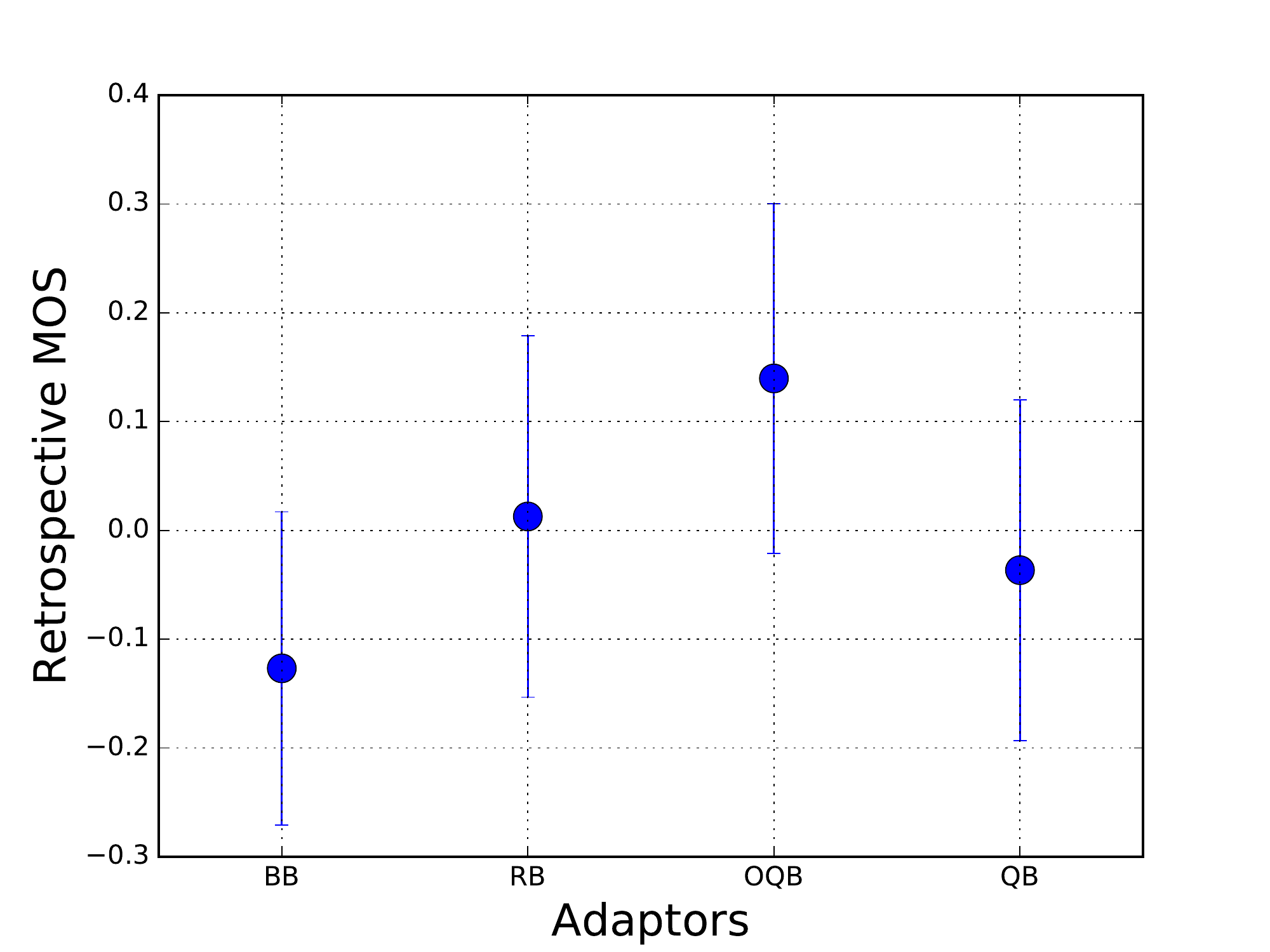}
}
\caption{Retrospective opinion score distribution for different adaptation algorithms (averaged across traces and contents). The error bars indicate the $95\%$ confidence interval.}
\label{FinalMOS_Adaptor_Analysis}
\end{figure}

As in the previous section, we compared the retrospective opinion scores among different adaptors (Fig. \ref{FinalMOS_Adaptor_Analysis}). We observed that the opinion scores are not very different across adaptors. This may be due to the fact that most of the rebuffering events occurred early in the video playout (as shown in Fig. \ref{RebufferingLocation_Analysis}), and because, just before the video finishes playing (and the retrospective score is recorded), the adaptation algorithms have built-up sufficient buffer to better handle bitrate/quality variations, even if the network is varying significantly. Therefore, it is likely that recency effects \cite{hands2001recency,bampis2017study} led to biases in the retrospective evaluations, i.e., subjects are forgiving/forgetful when recording retrospective QoE.

To validate this recency phenomenon, we averaged the continuous subjective scores over one second windows and calculated the correlation scores with the final subjective scores, as in \cite{bampis2017study}. For example, we found the continuous scores to correlate weakly in the $[4, 5]$ second window (correlation of $0.58$), while 20 seconds later, the correlation increased to $0.94$. Meanwhile, the per adaptor differences in terms of the average VMAF measurements were not considerably different, e.g., between RB and OQB, (see Table \ref{Objective_Aggregate_Analysis}) and hence the retrospective scores were also similar across adaptors.

\subsection{Analysis Using Continuous Scores}

\begin{figure}[htp]
\centerline{
\includegraphics[width=1\columnwidth]{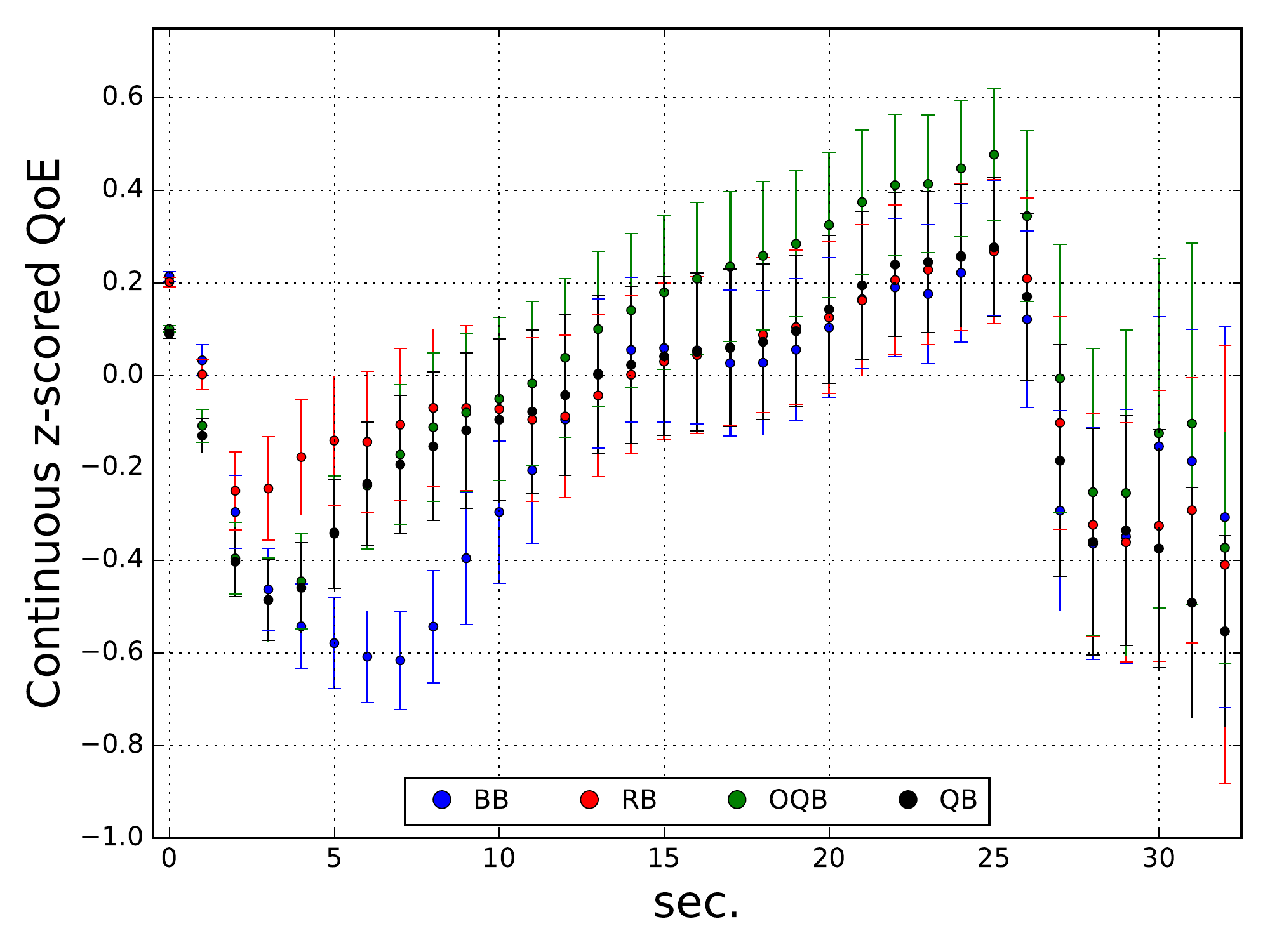}
}
\caption{Continuous-time scores for different adaptation algorithms (averaged across traces and contents). Error bars indicate $95\%$ confidence intervals.}
\label{ContinuousMOS_Adaptor_Analysis}
\end{figure}

Following our per-second objective analysis in Section \ref{objective_analysis}, Fig. \ref{ContinuousMOS_Adaptor_Analysis} depicts the continuous-time user experience across adaptation algorithms. We found that, within the first few seconds, the RB aggressive rate strategy initially leads to better QoE, unlike BB, QB and OQB, which opt for buffer build-up. This also means that subjects preferred increased early rebuffering, if it meant better start-up quality, as in the case of RB. Within the first 12 seconds, BB is overly conservative and delivers the lowest QoE among all adaptors, while QB and OQB perform between RB and BB. Nevertheless, after 12 seconds, QB and OQB improve considerably, with OQB tending to produce higher scores for the rest of the session. BB is relatively lower than RB and QB, both of which are statistically close. As before, we note that, after 25 seconds, QoE measurements are decreasing and have larger confidence intervals, since they correspond to videos that rebuffered, and their count decreases over time.

Notably, as in Fig. \ref{FinalMOS_Adaptor_Analysis}, we found that OQB is not statistically better than QB, even though it has perfect knowledge of the future bandwidth and performs the best in terms of objective metrics. As already explained, for the majority of distorted videos, rebuffering and quality degradations occurred earlier during video playout and this led to smaller differences in the subjective opinions per adaptor and over time. At this point, we want to clarify that this experimental result does not suggest that better bandwidth prediction is not an important goal, but shows that better bandwidth prediction does not significantly influence retrospective QoE scores. Meanwhile, the significant differences in QoE between adaptation strategies in the start-up phase underlines that temporal studies of QoE are highly relevant for adaptive video streaming, given that ABR algorithms are especially challenged during start-up.

\begin{figure}[tp]
\centerline{
\includegraphics[width=1\columnwidth]{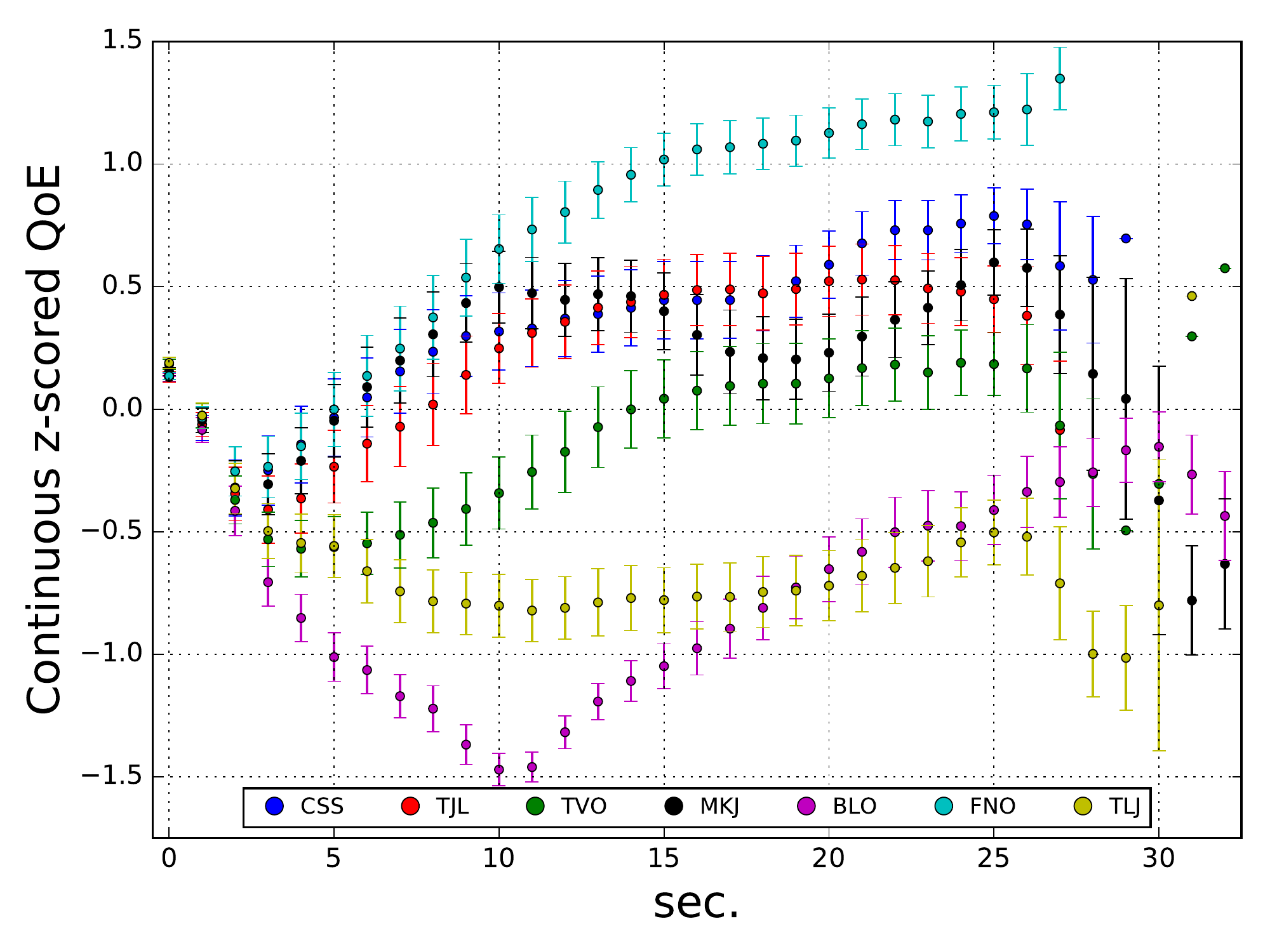}
}
\caption{Continuous-time scores for different network conditions. Error bars indicate $95\%$ confidence intervals.}
\label{ContinuousMOS_Trace_Analysis}
\end{figure}

Viewed from the network condition perspective, we found that continuous-time subjective scores are affected by dynamic quality/resolution changes and rebuffering. Figure \ref{ContinuousMOS_Trace_Analysis} shows that, for all traces, a few seconds are needed to build up video buffer and hence continuous scores are relatively low. Under better network conditions (e.g. FNO), user experience steadily improves after some time, due to the adaptors switching to higher resolution and lower compression ratio. By contrast, challenging cases such as BLO and TLJ recover slowly or do not recover at all, while very volatile conditions, as in MKJ, may also lead to noticeable drops in QoE much later during video playout. 

\subsection{Adaptation Algorithm Performance Discussion}

Following our earlier between-adaptor analysis, it is natural to ask which adaptation algorithm performs the best. In terms of retrospective scores, we were not able to make statistically significant comparisons, in part due to the effects of recency. On a similar note, using continuous-time scores, we found that OQB performed marginally better for a period of time, but the differences were not statistically significant even though OQB has perfect knowledge of the future bandwidth. By contrast, BB was overly conservative during startup and did not select high quality streams. 

Comparing RB and QB, we found that they delivered similar QoE over time, except during the start-up phase, where RB picked higher quality levels. The similar behavior between QB and RB can be attributed to their inherent properties: RB leads to excessive rebuffering, while QB reduces rebuffering (by taking into account the buffer level in its optimization scheme), but leads to many quality switches (see also Table \ref{Objective_Aggregate_Analysis}). In fact, an important consideration when designing QB is selection of the minimum buffer $B_l$ and target buffer $B_t$ values. When the network changes rapidly, the adaptor may not satisfy these and use its fallback mode, which leads to such large quality switches.

\subsection{Limitations of the LIVE-NFLX-II Database}

Despite our efforts in designing a diverse and realistic database that relies on state-of-the-art ideas in video encoding and streaming, one cannot overlook some remaining limitations. We recognise that QoE is not only affected by the factors investigated herein, such as visual quality, recency, rebuffering or quality switching, but also by other factors such as audio quality or contextual factors, like the display device and user expectations regarding the streaming service and/or the viewing environment. In our experiment, the audio quality was fixed and the display device was a computer monitor. 

Meanwhile, the adaptation algorithm design space and the number of possible network conditions are immense, hence our experiment can only capture the main characteristics of these dimensions as they pertain to user experience. Furthermore, the streaming sessions we generated are only between 25 and 36 seconds long (playback and rebuffering duration combined) and the network simulator does not consider the underlying TCP behavior, such as its slow restart property. Nevertheless, given the very large design space of the subjective experiment, it is virtually impossible to vary and explore all of the above streaming conditions simultaneously.

\section{Perceptual Video Quality and Quality of Experience}
\label{objective_VQA_QoE}

The perceptual optimization of adaptive video streaming requires accurate QoE prediction models \cite{mok2012qdash,rodriguez2012quality,xue2014assessing,chen2014modeling,liu2015deriving,bentaleb2016sdndash,duanmu2016sqi,robitza2017modular,raake2017bitstream,VATL,8315481}. These models either predict continuous-time QoE  or retrospective QoE. To form these predictions, video quality information is combined with other QoE indicators, such as the location and duration of rebuffering events, the effects of user memory, the effect of quality switching, and other factors.

An important goal of our database design is to use it as a development testbed for such video streaming quality and QoE prediction models. In this section, we evaluate a number of representative video quality assessment (VQA) and QoE prediction models. Given that the database contains both retrospective and continuous-time scores, we studied the performance of these algorithms both for retrospective and continuous-time QoE prediction applications. 

To calculate video quality, we decoded each distorted video into YUV420 format and applied each video quality model on the luminance channel of a distorted video and its reference counterpart. For video content with non-16:9 aspect ratio and, prior to VQA calculations, we also removed black bars to measure quality only for active pixels. For videos containing rebuffered frames, we removed all of those frames and calculated video quality on the aligned YUV files \cite{bampis2017study}. In the next sections, we investigate the predictive performance of leading VQA models and study their predictive performance when they are combined with QoE-driven models for retrospective and continuous-time QoE prediction.

\subsection{Objective Models for Retrospective QoE Prediction}

In our first experiment, we applied several well-known video quality and QoE metrics, including PSNR, PSNRhvs \cite{ponomarenko2007between}, SSIM \cite{wang2004image}, MS-SSIM \cite{wang2003multiscale}, ST-RRED \cite{soundararajan2013video}, VMAF \cite{techblog} (version 0.6.1), SQI \cite{duanmu2016sqi} and Video ATLAS \cite{VATL}. The original Video ATLAS model \cite{VATL}, was designed and tested on the LIVE-NFLX and Waterloo databases, where spatial resolution changes and quality switching events were much less diverse. Given the flexibility of Video ATLAS and the diversity of our newly designed database, we re-trained the model to include changes on resolution and quality. 

\begin{figure}[tp]
\centerline{
\includegraphics[width=0.8\columnwidth]{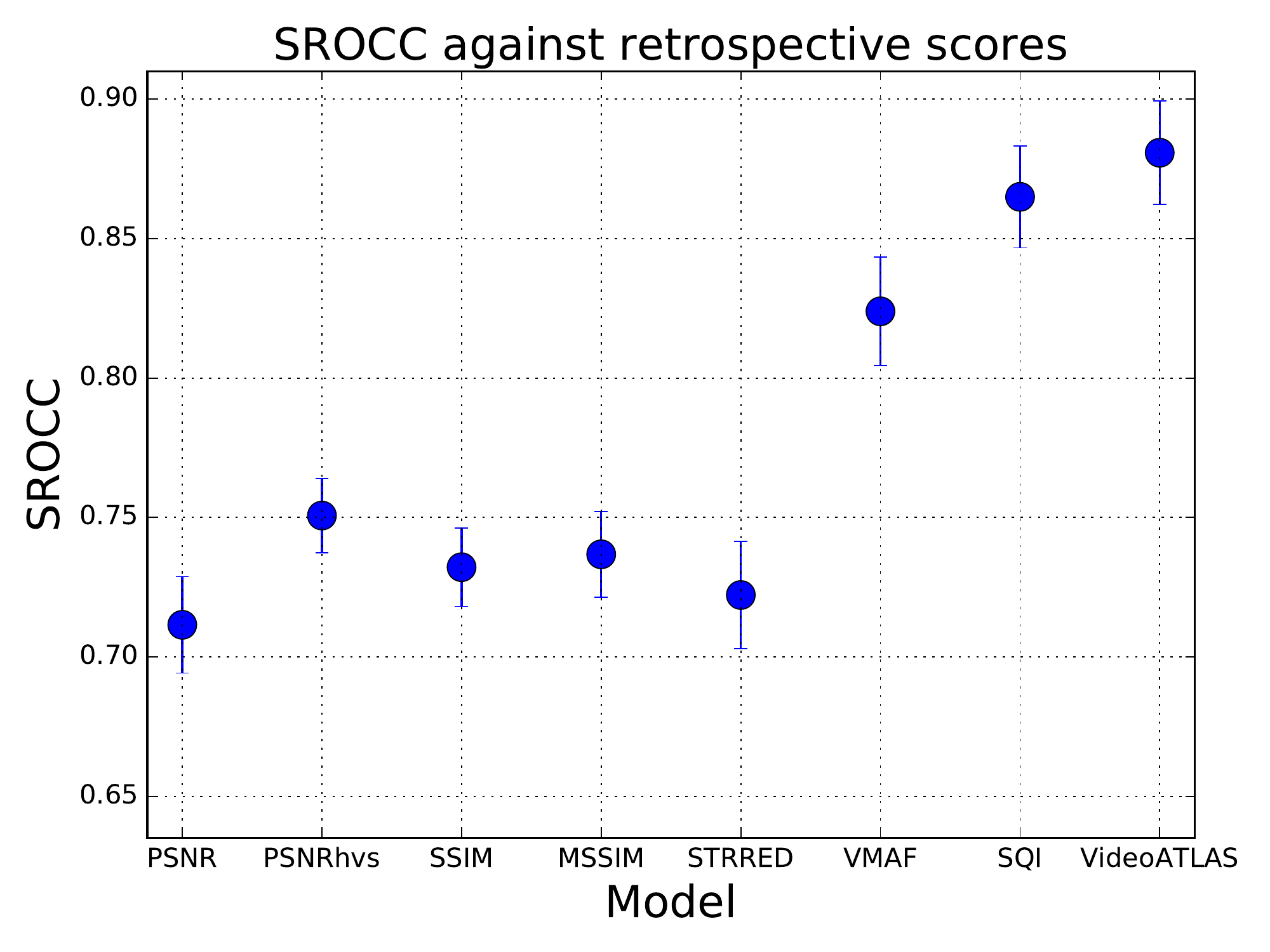}
}
\caption{Boxplots of SROCC performance of leading VQA and QoE models using retrospective scores. The error bars indicate the $95\%$ confidence interval.}
\label{Boxplot_retrospective}
\end{figure}

We used VMAF as the VQA feature, average absolute difference of encoding resolution (to capture the effects of resolution switching) and rebuffer duration as features. We also kept track of the time in seconds (from the end of the video) since the most recent minimum quality was observed (TLL feature). For SQI, VMAF was also used as the VQA model. We excluded the P.1201-3 models \cite{stand}, since they are trained for video sequences longer than one minute. To evaluate performance, we use Spearman's Rank Order Correlation Coefficient (SROCC), which measures the monotonicity between groundtruth QoE and predictions. The results of our experiment are shown in Figure \ref{Boxplot_retrospective}.

Since Video ATLAS is a learning-based model, we split the database into multiple train/test splits. When using image and video quality databases, it is common to split the database into content-independent splits; but for the streaming scenario we propose a different approach. Given that video contents are pre-encoded and the behavior of an adaptation algorithm is deterministic (given a network trace and a video content), it is more realistic to assume that, during training, we have collected subjective scores on a subset of the network traces. Therefore, we perform our splitting based on the network traces by choosing 5 traces for training and 2 for testing each time, which yields ${7 \choose 2}=21$ unique combinations of 300 (15 contents, 4 adaptors and 5 traces) training and 120 (15 contents, 4 adaptors and 2 traces) testing videos. The total number of combinations may not be as large; but each train/test subset contains hundreds of videos. Figure \ref{Boxplot_retrospective} shows boxplots of performance across all 21 iterations for all compared models.

It can be observed that all of the VQA-only models lagged in performance, which is to be expected since VQA models only capture visual quality and disregard other critical aspects of QoE such as rebuffering. Nevertheless, VMAF performed significantly better than all other models. As a reminder, VMAF was used on multiple occasions when generating the final videos, e.g., when generating the bitrate ladder, deciding on encoding parameters and performing client-based adaptation for QB and OQB. This suggests that our system is better tuned towards the VMAF model and that the choice of quality model has a direct impact on user experience. Using VMAF as part of the SQI and Video ATLAS QoE predictors led to performance gains in both cases.

To validate the contribution of VMAF in predicting QoE, and measure the contribution of the other three factors in Video ATLAS (rebuffering duration, resolution switching and TLL), we also trained a simple Random Forest (RF) regressor and measured feature importance over all 21 unique train/test combinations. The number of estimators in RF was set to 100 and the feature importances were found to be: 79\% for VMAF, 11\% for rebuffering duration, 5\% for resolution switching and 5\% for TLL. This validates the observation that VMAF scores contribute strongly to QoE prediction, while rebuffering duration is less important, in part because rebuffering events occurred earlier in the streaming session and hence were less important for retrospective QoE evaluations. We tested a wide range of RF estimators ($10...1000$) and got similar results.

\subsection{Objective Models for Continuous-time QoE Prediction}

Predicting continuous-time QoE is a harder task, given the challenges in collecting reliable ground truth data and designing models that can integrate perceptually-motivated properties into a time-series prediction. Earlier approaches \cite{chen2014modeling} have addressed the problem of predicting time-varying quality and only recently similar works have addressed time-varying QoE prediction \cite{8315481}. 

To evaluate performance, we used root mean squared error (RMSE) and outage rate (OR). RMSE measures the prediction's fidelity to the ground truth, while OR measures the frequency of predictions falling outside twice the confidence interval of subjective scores.

We evaluated two prediction algorithms presented in \cite{8315481}, one based on autoregressive neural networks (G-NARX) and the other based on recurrent neural networks (G-RNN). We note that using the SQI model in \cite{duanmu2016sqi} to predict continuous-time QoE does not deliver scores that fall within the z-scored continuous MOS scale, since it is not trained on subjective data. Therefore, the RMSE and OR values are considerably worse. Further, using SROCC as an evaluation metric did not yield satisfactory results (an SROCC value of approximately 0.41) and using SROCC may not be an appropriate choice for comparing time-series \cite{8315481}.

To train the G-NARX and G-RNN models, we used per-frame VMAF measurements as the continuous-time VQA feature. We also included two more continuous-time features: a per-frame boolean variable denoting the presence of rebuffering and another denoting the time since the latest rebuffer. We used 8 input delays and 8 feedback delays for G-NARX and 5 layer delays for G-RNN. Both approaches used 8 hidden nodes and the training process was repeated three times yielding an ensemble of three test predictions per distorted video that were averaged for more reliable time-series forecasting. We configured the prediction models to output one value per 0.25 sec., by averaging the continuous-time variables accordingly.

\begin{table}[htp]
\caption{Prediction performance of the G-NARX and G-RNN QoE models using continuous scores.}
\centering
\scalebox{1}{
\begin{tabular}{| c | c | c |}
\hline
Model & RMSE & OR \\ \hline
G-NARX & 0.267 & 7.136\% \\ \hline
G-RNN & 0.276 & 5.962\% \\ \hline
\end{tabular}}
\label{ContinuousResults}
\end{table}

\begin{figure}[tp]
\centerline{
\includegraphics[width=0.9\columnwidth]{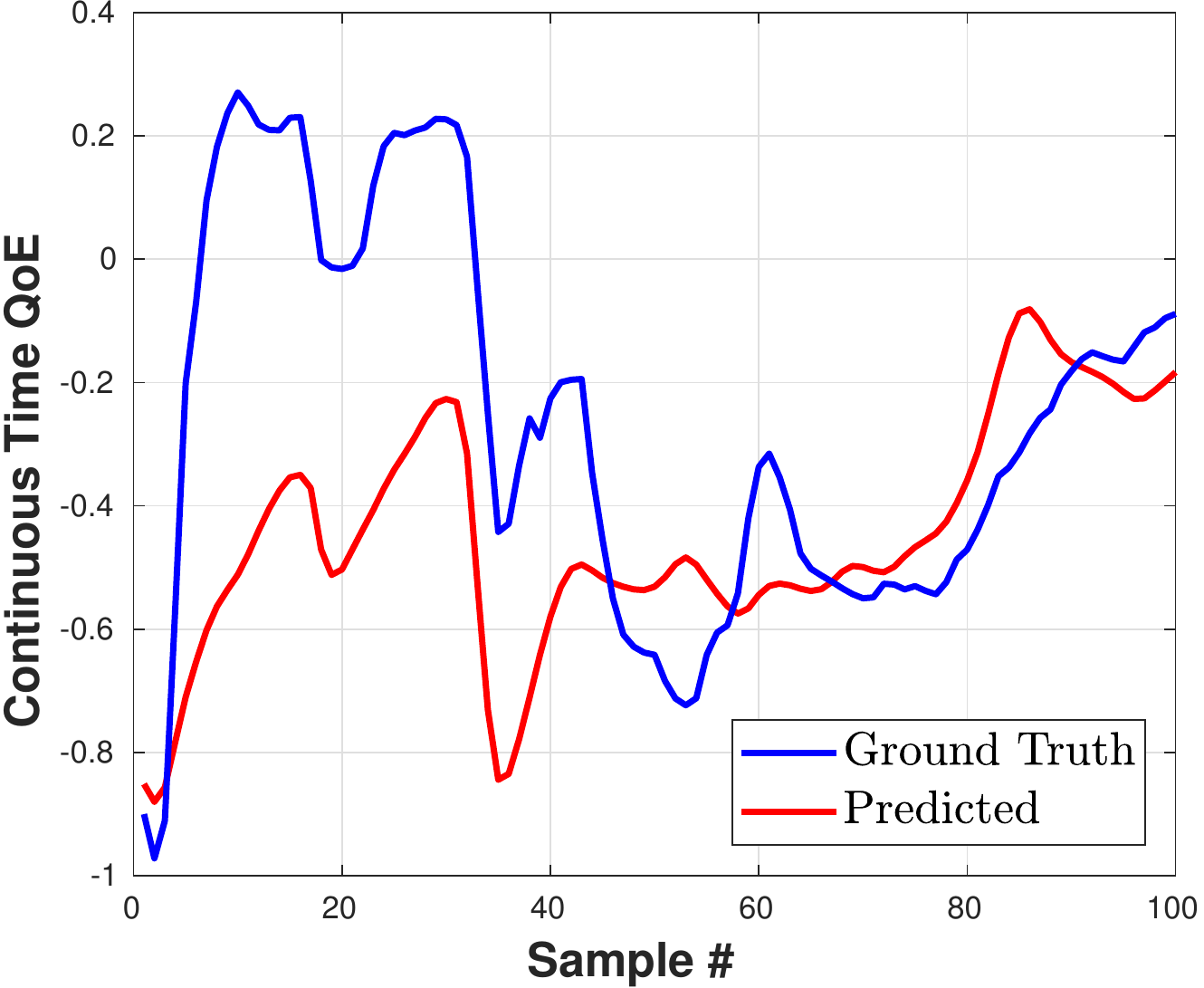}
}
\caption{An example where the G-NARX QoE prediction does not capture trends in subjective QoE.}
\label{cont_cases}
\end{figure}

Table \ref{ContinuousResults} shows that both approaches delivered promising performance and similar to each other in terms of RMSE and OR. Nevertheless, we observed cases where predictions could be further improved, as in Fig. \ref{cont_cases}. In this case, the G-NARX QoE prediction did not accurately capture trends in subjective QoE and its dynamic. In similar test results, we identified that perhaps the main problem of these trained networks is that they do not always capture the magnitude of subjective opinion changes over time, i.e. they tend to over- or under-estimate a drop or an increase in QoE. This demonstrates the need to integrate better human perception models, to accurately capture continuous QoE responses.

\section{Discussion and Conclusion}
\label{the_end}

We presented the design of a large, comprehensive subjective video database, which relied on a highly realistic streaming system. The collected data allowed us to analyze overall and continuous-time user experiences under different network conditions, adaptation algorithms and video contents. Using the collected human opinion scores, we also trained and evaluated predictors of video quality and quality of experience. 

In the future, we intend to use the ground truth data to build better continuous-time QoE predictors by integrating additional features, such as resolution changes, network estimates and buffer status. Inspired by similar QoE works as in \cite{joseph2014nova}, our ultimate goal is to ``close the loop", i.e., inject such QoE-aware predictions into the client-adaptation strategy in order to perceptually optimize video streaming.

\section{Acknowledgement}
The authors thank Anush K. Moorthy for discussions regarding content encoding complexity and the content-adaptive bitrate ladder. Also, Anne Aaron and the entire Video Algorithms team at Netflix for supporting this work.

\section{Appendix: Encoding Module, Video Quality Module and the Streaming Pipeline}

\subsection{Video Encoding}
\label{encoding_module_appendix}

\begin{figure}[htp]
\centerline{
\includegraphics[width=0.8\columnwidth]{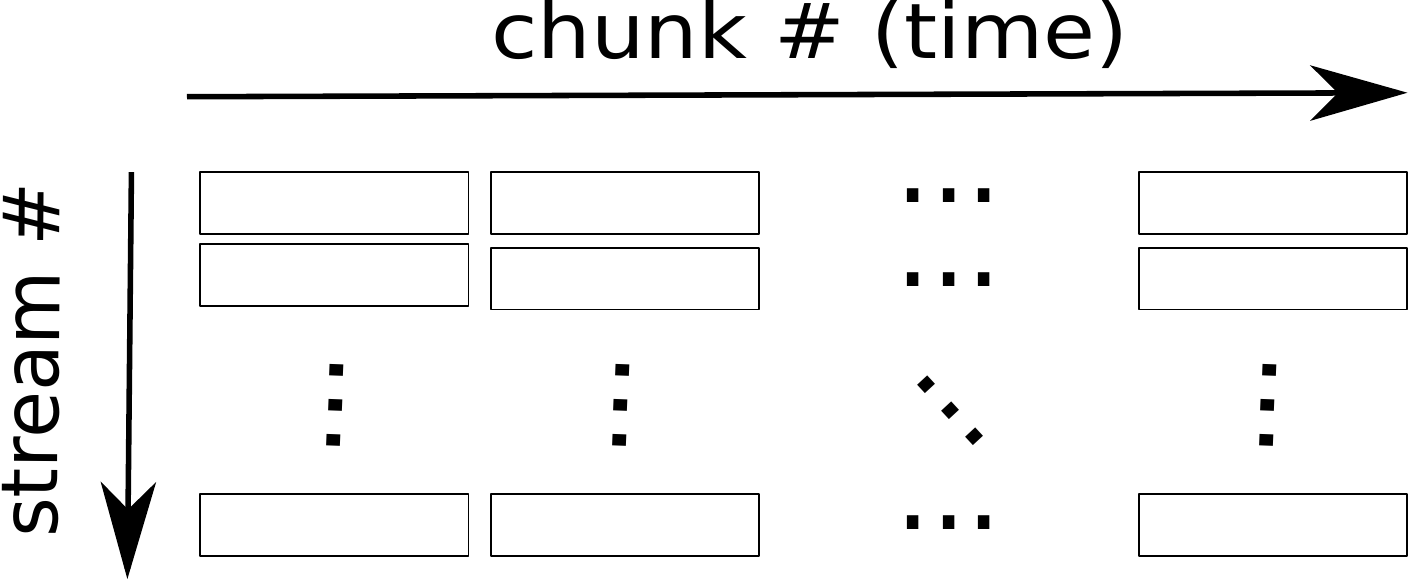}
}
\caption{An encoding chunk map representation.}
\label{encoding_chunk_map}
\end{figure}

To generate the video encodes, we adopted the Dynamic Optimizer (DO) \cite{DO_techblog} approach that was recently developed and implemented by Netflix. DO determines the optimal encoding parameters per shot, such that a pre-defined metric is optimized at a given bitrate. The underlying assumption is that video frames within a video shot have similar spatio-temporal characteristics (e.g. camera motion and/or spatial activity), and hence should be encoded at a particular resolution and QP value. For a specific target bitrate, the DO implementation determines the ``optimal" resolution and QP values per shot that meets (but does not exceed) this bitrate while maximizing the overall quality predicted by the VMAF model \cite{techblog}. Repeating this process over each target bitrate value and video segment yields a 2D encoding chunk map, where each row is a single video stream (corresponding to the same target bitrate) and each column is a different video segment over time (see Fig. \ref{encoding_chunk_map}) encoded at the QP and resolution values selected by the DO.



Both the encoding and client modules are driven by VMAF \cite{techblog} measurements which are carried out by the video quality module. In the following section, we discuss the video quality module in detail.

\subsection{Video Quality}
\label{video_quality_module_appendix}

The end receiver of any video streaming service is the human eye, hence integrating video quality models in a streaming network is of paramount importance to achieve perceptually optimized video streaming. Numerous studies have shown that simple video quality indicators such as the encoding bitrate or peak signal-to-noise ratio (PSNR) do not correlate well with human perception \cite{wang2004image,wang2009mean}. Nevertheless, PSNR is still often used for codec optimization and codec comparisons \cite{techblog}, while encoding bitrates are widely used by client adaptation algorithms.

In our model system, the video quality module performs perceptual video quality calculations that are fed to the encoding module, in order to determine an appropriate bitrate ladder (the set of target bitrates per content), construct its convex hull \cite{de2016complexity,DO_techblog} and determine the encoding parameters. The same video quality module is also fed to the client module in order to drive quality-based streaming decisions. We also use video quality measurements to perform offline analysis of the final video sequences and to compare them with human subjective scores (see Section \ref{objective_VQA_QoE}). To effectively measure quality, we relied on the VMAF model \cite{techblog}. The choice of VMAF is not restrictive, and other high-performing video quality indicators can also be used.

VMAF is used to encode and monitor the quality of millions of encodes on a daily basis \cite{techblog} and exhibits a number of key properties. It has been trained on streaming-related video impairments, such as compression. Its elementary features, such as the VIF model \cite{sheikh2006image,sheikh2005visual}, are highly descriptive of perceptual quality. It is also more computationally efficient than time-consuming VQA models such as MOVIE \cite{seshadrinathan2010motion} and VQM-VFD \cite{pinson2014temporal}. Further, the VMAF framework is publicly available \cite{vmaf_github} and can be improved even further by adding other quality-aware features or regression models. Lastly, given that it is a trained algorithm, it produces scores that are linear with the subjective scale, i.e., a VMAF of 80 (VMAF $\in [0, ..., 100]$) means that, on average, viewers will rate a video with a score of 8 out of 10.

\subsection{Putting the Pieces Together}
\label{pieces_together_appendix}

\begin{figure*}[tp]
\centerline{
\includegraphics[width=1.6\columnwidth]{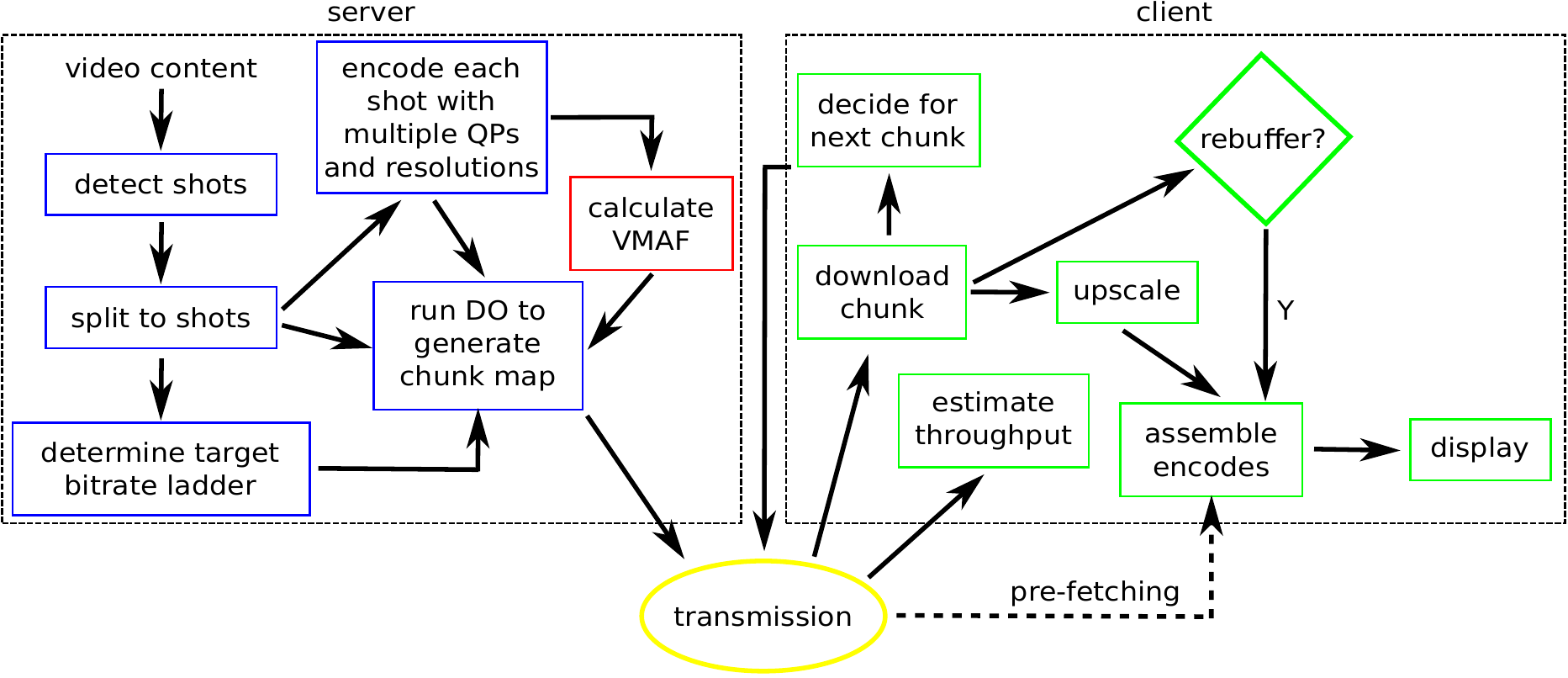}
}
\caption{Full pipeline of LIVE-NFLX-II, where each module is color-coded: blue: encoding module; red: video quality module, yellow: network module and green: client module. The client's behavior is orthogonal to the offline video encoding and quality calculations on the server side.}
\label{full_pipeline}
\end{figure*}

After having described most of the individual components, we can give an overview of the end-to-end streaming pipeline, as depicted in Fig. \ref{full_pipeline}. First, the encoding module performs shot detection and splits the video content into different shots \cite{DO_techblog}. Each of these shots is then encoded at multiple encoding levels, each determined by an encoding resolution and a QP value. The video quality module calculates the average per segment VMAF values for each of the pre-encoded segments. After determining the target bitrates for each content, these bitrates are then passed (along with VMAF values and encoded chunks) into the DO, which decides on the per shot encoding resolution and QP values. This results in a 2D encoding chunk map (also depicted in Fig. \ref{encoding_chunk_map}). Notably, these steps are carried out in an offline fashion and are orthogonal to the client's behavior and/or the network condition.

To exclude the effects of pre-buffering on our analysis, the client algorithm first pre-fetches a $B_0=1$ chunk. This pre-fetched chunk was always encoded at the lowest bitrate/quality. Based on the client algorithm (BB, RB, QB or OQB), the client then decides which stream should be selected for the next chunk. If the buffer is depleted, a rebuffering event occurs. To simulate rebuffering, we retrieve the latest frame that was played out, and overlay a spinning wheel icon on the viewing screen. Rebuffering occurs until the buffer is sufficiently filled to display the next chunk and the client adaptation algorithm allows for playback to resume. Before display, each encode is upscaled using bicubic interpolation, to match the 1080p display resolution. In the case of QB and OQB, the client uses the VMAF values of future segments (within the horizon $h$) to drive its decision. Note that for RB and QB, the available throughput is estimated by averaging download speeds of the most recent $w=5$ chunks.

\subsection{Detailed Data About Network Traces}
\label{network_traces_analytic_table_appendix}

In this brief Section, we provide more details on the network traces in Table \ref{network_traces_analytic_table}. It may be observed that there are multiple types of transportation and multiple routes included in our real traces. In terms of the available bandwidth range, the traces cover the range of 9 Kbps up to almost 3900 Kbps.

\begin{table}[htp]
\caption{Summary of the network traces used in LIVE-NFLX-II. The available bandwidth $B$ is reported in Kbps. We denote by min $B$, max $B$, $\mu_B$ and $\sigma_B$ the minimum, maximum, average and standard deviation of the available bandwidth.}
\centering
\scalebox{0.7}{
\begin{tabular}{| c | c | c | c | c | c | c | c |}
\hline
ID & Type & min $B$ & max $B$ & $\mu_B$ & $\sigma_B$ & From & To \\ \hline
CSS & Car & 234 & 1768 & 989 & 380 & Snaroya & Smestad \\ \hline
TJL & Tram & 52 & 1067 & 617 & 207 & Jernbanetorget & Ljabru \\ \hline
TVO & Train & 131 & 1632 & 702 & 349 & Vestby & Oslo \\ \hline
MKJ & Metro & 28 & 1511 & 696 & 456 & Kalbakken & Jernbanetorget \\ \hline
BLO & Bus & 9 & 886 & 373 & 235 & Ljansbakken & Oslo \\ \hline
FNO & Ferry & 35 & 3869 & 1325 & 761 & Nesoddtangen & Oslo \\ \hline
TLJ & Tram & 86 & 485 & 269 & 86 & Ljabru & Jernbanetorget \\ \hline
\end{tabular}}
\label{network_traces_analytic_table}
\end{table}

\bibliographystyle{IEEEtran}
\bibliography{live_nflx_plus_journal_bib}{}

\end{document}